%% file: template.tex
\documentclass[preprint]{vgtc}               

\ifpdf
  \pdfoutput=1\relax                   
  \usepackage{graphicx}                
  \DeclareGraphicsExtensions{.pdf,.png,.jpg,.jpeg} 
\else
  \usepackage{graphicx}                
  \DeclareGraphicsExtensions{.eps}     
\fi%

\graphicspath{{figures/}{pictures/}{images/}{./}} 

\input{Additional_Packages.tex}

\usepackage{microtype}                 
\PassOptionsToPackage{warn}{textcomp}  
\usepackage{textcomp}                  
\usepackage{mathptmx}                  
\usepackage{times}                     
\usepackage{cite}                      
\usepackage{tabu}                      
\usepackage{booktabs}                  

\onlineid{0}

\vgtccategory{Research}

\vgtcinsertpkg

\preprinttext{To appear in an IEEE VGTC sponsored conference.}

\usepackage[nameinlink,capitalise]{cleveref}
  \crefname{table}{Tab.}{Tabs.}
  \Crefname{table}{Table}{Tables}
  \crefname{section}{Sec.}{Secs.}
  \Crefname{section}{Section}{Sections}

\title{A Mathematical Foundation for the Spatial Uncertainty\texorpdfstring{\\}{ }of Critical Points in Probabilistic Scalar Fields}

\author{Dominik Vietinghoff\thanks{e-mail: vietinghoff@informatik.uni-leipzig.de}\\ %
    \footnotesize Leipzig University %
\and Michael B\"ottinger\thanks{e-mail: boettinger@dkrz.de}\\ %
    \footnotesize Deutsches Klimarechenzentrum 
\and Gerik Scheuermann\thanks{e-mail: scheuermann@informatik.uni-leipzig.de}\\ %
    \footnotesize Leipzig University %
\and Christian Heine\thanks{e-mail: heine@informatik.uni-leipzig.de}\\ %
    \footnotesize Leipzig University %
}

\abstract{ %
    Critical points mark locations in the domain where the level-set topology of a scalar function undergoes fundamental changes and thus indicate potentially interesting features in the data. Established methods exist to locate and relate such points in a deterministic setting, but it is less well understood how the concept of critical points can be extended to the analysis of uncertain data. Most methods for this task aim at finding likely locations of critical points or estimate the probability of their occurrence locally but do not indicate if critical points at potentially different locations in different realizations of a stochastic process are manifestations of the same feature, which is required to characterize the spatial uncertainty of critical points. Previous work on relating critical points across different realizations reported challenges for interpreting the resulting spatial distribution of critical points but did not investigate the causes. In this work, we provide a mathematical formulation of the problem of finding critical points with spatial uncertainty and computing their spatial distribution, which leads us to the notion of uncertain critical points. We analyze the theoretical properties of these structures and highlight connections to existing works for special classes of uncertain fields. We derive conditions under which well-interpretable results can be obtained and discuss the implications of those restrictions for the field of visualization. We demonstrate that the discussed limitations are not purely academic but also arise in real-world data.%
} 

\CCScatlist{Uncertainty visualization, scalar field topology, critical points, ensemble data.}

\begin{document}

\setlength{\abovedisplayskip}{4pt}
\setlength{\belowdisplayskip}{4pt}

\setlength{\abovecaptionskip}{3pt}
\setlength{\belowcaptionskip}{-16pt}


\firstsection{Introduction\label{sec:introduction}}

\maketitle

\input{sections/Introduction}
\input{sections/RelatedWork.tex}
\input{sections/Background.tex}
\input{sections/Methods.tex}
\input{sections/Conclusion.tex}

\acknowledgments{%
This work was funded by the Deutsche Forschungsgemeinschaft (DFG, German
Research Foundation) - SCHE 663/11-2.
}

\bibliographystyle{abbrv-doi-hyperref}

\bibliography{template}
\end{document}

%% file: Additional_Packages.tex
\usepackage{amsmath}
\usepackage{amsfonts}
\usepackage{amssymb}
\usepackage{dsfont}
\usepackage[per-mode=symbol]{siunitx}
\usepackage[svgnames]{xcolor}
\usepackage{tikz}
\usetikzlibrary{positioning, calc}
\usepackage{pstricks}
\usepackage{pstricks-add}
\usepackage[shortlabels]{enumitem}

\usepackage{comment}

\makeatletter
\newcommand{\creffigname}{\cref@figure@name}
\newcommand{\Creffigname}{\Cref@figure@name}
\makeatother
\newcommand{\crefSubFigRef}[2]{%
    \crefformat{figure}{##2\creffigname{}~##1{#2}##3}%
    \cref{#1}%
    \crefformat{figure}{##2\creffigname{}~##1##3}%
}
\newcommand{\CrefSubFigRef}[2]{%
    \Crefformat{figure}{##2\Creffigname{}~##1{#2}##3}%
    \Cref{#1}%
    \Crefformat{figure}{##2\Creffigname{}~##1##3}%
}

\usepackage{amsthm}
\usepackage{thmtools}
\declaretheoremstyle[
    qed=$\diamond$, 
    spaceabove=2pt,
    spacebelow=2pt
]{theorem-style}
\declaretheorem[style=theorem-style]{lemma}
\declaretheorem[style=theorem-style, sibling=lemma]{proposition}
\declaretheorem[style=theorem-style, sibling=lemma]{remark}
\declaretheorem[style=theorem-style, sibling=lemma]{definition}
\declaretheorem[style=theorem-style, sibling=lemma]{example}

\declaretheoremstyle[
    spaceabove=2pt, 
    spacebelow=2pt, 
    headfont=\normalfont\itshape, 
    bodyfont = \normalfont,
    postheadspace=1em, 
    qed=$\square$, 
]{proofstyle} 
\declaretheorem[name={Proof}, style=proofstyle, unnumbered]{Proof}

\newcommand{\etal}{et al.}

\DeclareMathOperator{\rank}{rank}
\DeclareMathOperator{\sgn}{sgn}

\newcommand{\domainDims}{d}
\newcommand{\numGridPts}{n}
\newcommand{\numFields}{m}
\renewcommand{\vec}[1]{\mathbf{#1}}
\newcommand{\mat}[1]{\mathbf{#1}}

\newcommand{\domJacobi}{\mathcal{M}}
\newcommand{\ndCritPts}{\mathsf{M}}
\newcommand{\dom}{\mathcal{D}}
\newcommand{\grid}{\mathcal{G}}
\newcommand{\gridPt}{x}
\newcommand{\param}{a}
\newcommand{\rvParam}{A}
\newcommand{\gradMat}{\mat{G}}

\newcommand{\reals}{\mathds{R}}
\newcommand{\prob}{\mathds{P}}
\newcommand{\realSpace}{\mathcal{R}}
\newcommand{\paramSpace}{\mathcal{P}}
\newcommand{\constraint}{\mathcal{C}}

\newcommand{\sameness}{\equiv}
\newcommand{\fun}{f}
\newcommand{\evs}{g}
\newcommand{\helper}{h}
\newcommand{\pdf}{\phi}
\newcommand{\cdf}{\Phi}
\newcommand{\ucp}{\mathsf{C}}

\newcommand{\jacobi}{\mathcal{J}}
\newcommand{\jacobiDiscr}{\mathds{J}}

%% file: sections/Introduction.tex
A well-known summarization approach in visualization of deterministic fields is to highlight topological features in the data, but research extending these concepts to ensembles of fields, let alone probability distributions over fields, is still in its early stages~\cite{Heine2016}.
In this work, we focus on critical points, a fundamental concept and building block in the topology of scalar fields.
To avoid manual inspection of a large number of ensemble members, visual summaries that assemble the information about critical points in a single image are desirable.
Previous work has aimed to characterize regions in the domain that with high certainty contain a critical point~\cite{Mihai2014, Guenther2014, Vietinghoff2021} or estimate the probability of a point in the domain being critical~\cite{Vietinghoff2022, Vietinghoff2022a}.

\begin{figure}
    \centering
    \begin{tikzpicture}
    \node[inner sep=0pt] (coarse) {%
        \includegraphics[width=.25\linewidth]{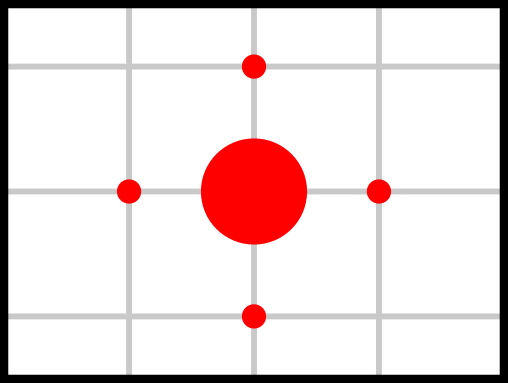}
    };
    \node[inner sep=0pt, right=2em of coarse] (dense) {%
        \includegraphics[width=.25\linewidth]{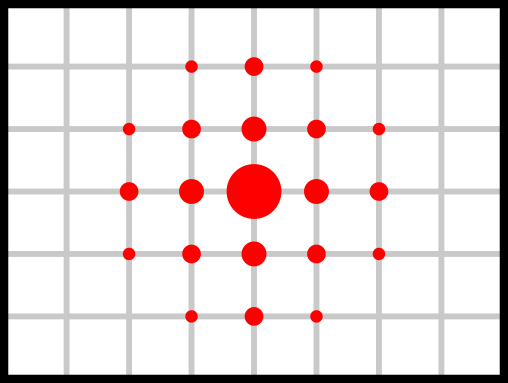}
    };
     \node[inner sep=0pt, right=2em of dense] (density) {%
        \includegraphics[width=.25\linewidth]{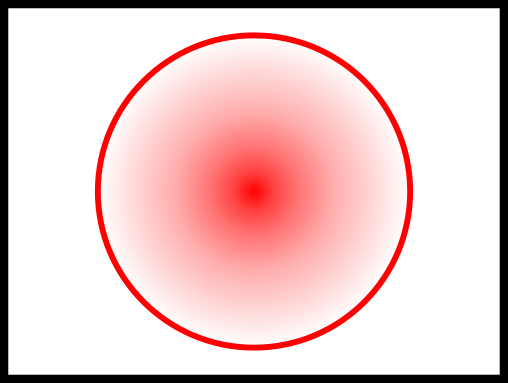}
    };
    \end{tikzpicture}
    \caption{Shortcoming of local approaches for extending the concept of critical points to uncertain data. The high probability (large glyph) for a critical point on a coarse grid (left) is distributed over multiple grid points in a finer resolved data set (middle) indicating spatial uncertainty of the critical point. Right: An intuitive representation of a critical point with spatial uncertainty.}
    \label{fig:shortcomings-local-methods}
\end{figure}

Mihai and Westermann~\cite{Mihai2014} investigate how data variation propagates through finite differences approximations of the gradient and Hessian matrix to find points that are likely to be critical and decide on the most probable type of the critical point. However, this only yields a binary classification of each point and no quantitative measure of the likelihood of a critical point to occur.
Vietinghoff \etal{} presented a Bayesian approach for the computation of occurrence probabilities of critical points~\cite{Vietinghoff2022} and confidence intervals for those probabilities~\cite{Vietinghoff2022a} in ensemble data sets to communicate epistemic uncertainties that may arise from (too) few ensemble members. Both methods operate locally on a per-grid-vertex basis. 
With rising grid resolution, the uncertainty in the data is likely to cause critical points to occur at vertices with slightly different locations causing the probability to observe a critical point at a specific vertex to converge to zero (see \cref{fig:shortcomings-local-methods} left and middle). As a consequence, the outputs of those methods would become meaningless. 
This effect of a likely critical point in a sparsely sampled representation of the data becoming multiple less likely critical points in a higher resolved version indicates spatial uncertainty of the underlying topological feature.

This raises the question of how to characterize critical points with spatial uncertainty. Intuitively one would imagine a connected spatial region encompassing all locations where the critical point might occur and a spatial probability distribution over this region describing how likely it is to observe the critical point at any given location (see \cref{fig:shortcomings-local-methods} right). Because all three methods discussed above \cite{Mihai2014,Vietinghoff2022,Vietinghoff2022a} decide only locally on the likelihood of a point to be critical and do not relate the occurrence of critical points at neighboring grid points, they are inherently unable to provide this kind of spatial information. A more promising approach in this direction are \emph{mandatory critical points} introduced by Günther \etal~\cite{Guenther2014}, which are regions in the domain where every realization of an uncertain scalar field must have at least one critical point of a given type. By construction, this ignores regions where some or even a majority of realizations, yet not all, have a critical point. In practice, however, it might be those features that occur only under special circumstances that are most interesting to the domain scientists.

Liebmann and Scheuermann~\cite{Liebmann2016} presented a different approach to this question in the context of Gaussian-distributed piecewise-linear (PL) scalar fields. They compute so-called \emph{singular patches} that describe subsets of realizations that attain a critical point of a certain type at some grid point. On those patches, they introduce an adjacency relation, which allows them to merge singular patches of neighboring grid points to capture the spatial uncertainty of a critical point. They found, however, that if not stopped prematurely, this merging process may produce counterintuitive results of the same critical points occurring at different locations in the same field and spatial probability distributions that are difficult to interpret.
Because their method is motivated through combinatorial arguments based on the discrete representation of the data rather than an analysis of the underlying mathematical problem, we wondered if the observed issues are specific to Liebmann and Scheuermann's algorithm, the used PL representation of the uncertain scalar field, or if there is a more fundamental problem preventing an intuitive definition of critical points with spatial uncertainty. In this paper, we therefore analyze this problem from a more theoretical perspective.

While our work was initially motivated by the wish to overcome the issues discovered by Liebmann and Scheuermann, ultimately, this paper does not provide an (algorithmic) solution to those problems. In fact, our research revealed (see \cref{example:counterexample} in \cref{sec:relation-to-jacobi-sets}) that a simple extension of critical points to uncertain data with the goal of capturing the spatial distribution of a critical point might be impossible with our current understanding of what a critical point with spatial uncertainty should look like, thus supporting the statement by Heine et al.~\cite{Heine2016} that we might require entirely new descriptors for the topology of uncertain scalar fields.
In summary, the contributions of this paper are threefold:
\begin{enumerate}[noitemsep,topsep=0pt,leftmargin=1.4em]
    \item We provide a rigorous discussion on the concept of critical points with spatial uncertainties in a generic setting (\cref{sec:theoretical-considerations}). For this, we analyze if and when critical points in different realizations of an uncertain scalar field describe the same topological feature, which leads to the notion of \emph{uncertain critical points} (\cref{sec:uncertain-critical-points}). We also describe how to extract information on the spatial distribution of a critical point from this concept (\cref{sec:spatial-distribution}).
    \item We explain how the concept of uncertain critical points is connected to Jacobi sets~\cite{Edelsbrunner2004} in the case that the realization space can be parameterized with a finite set of parameters (\cref{sec:relation-to-jacobi-sets}). In this context, we also discuss conditions under which the issues described by Liebmann and Scheuermann do not arise and provide a synthetic example (\cref{example:counterexample}) demonstrating that those conditions are sharp and highlighting the severity of the problem.
    \item We discuss how the work from Liebmann and Scheuermann~\cite{Liebmann2016} and previous works on the temporal tracking of critical points~\cite{Edelsbrunner2008, Mascarenhas2006} in PL data fit the description of uncertain critical points with the Jacobi set (\cref{sec:algo-discr}). That is, we show that our concepts can be seen as a generalization of those combinatorial approaches. 
\end{enumerate}

%% file: sections/RelatedWork.tex
\section{Related Work}
\label{sec:relatedwork}
Johnson~\cite{Johnson2004} identified the visualization of uncertain data as one of the top scientific visualization research problems.
The advances in the field of uncertainty visualization since then have been summarized in several state-of-the-art reports~\cite{Bonneau2014, Brodlie2012, Potter2012, Wang2019, Kamal2021}. However, in their survey on topology-based methods in visualization, Heine \etal~\cite{Heine2016} identified a lack of methods that deal with uncertain data. In addition to the works on critical point analysis on uncertain fields mentioned in \cref{sec:introduction}, this section outlines further related works.

\noindent\textbf{Topology of Uncertain Scalar Fields.}
For deterministic fields, contour (CTs) and merge trees (MTs) are popular topology descriptors describing the nesting properties of the components of a field's (super- and sub-)level sets. Some approaches have been made to extend these structures to uncertain data. Kraus~\cite{Kraus2010} describes an algorithm for merging two CTs where one is a pruned version of the other. He then visualizes the merged tree and highlights parts where the two input trees differ. Wu and Zhang~\cite{Wu2013} present three ways to encode uncertainties in visual representations of CTs that arise from variations on the data, contour, and topology levels.
 Zhang \etal~\cite{Zhang2015} provide data structures and algorithms to compute the probability of two domain points to be mapped to the same CT edge and to compute the expected distance of such points within the CT. 
 Lohfink \etal~\cite{Lohfink2020} compute an alignment tree, containing each CT of an ensemble of input fields as a subtree. 
 In a planar drawing of this tree, they group aligned branches of the input trees and use edge bundling to reduce visual clutter, resulting in a \emph{fuzzy contour tree}.

Similarly, Yan \etal~\cite{Yan2020} compute the \emph{1-center} of a set of MTs and visualize the \emph{vertex consistency} of the input trees and the average tree.
Sridharamurthy \etal~\cite{Sridharamurthy2020} measure the similarity of MTs with an edit distance. 
Extending this distance, Pont \etal~\cite{Pont2021} define the \emph{Wasserstein distance between MTs} and 
use it to compute geodesics between two MTs and the barycenters of a set of MTs. 
Recently, Pont \etal~\cite{Pont2023} built on this work to introduce the notion of a \emph{principal geodesic analysis of MTs}.

All those methods operate by combining or interpolating the analyzed topological descriptors of a finite set of fields, which requires matching features of different fields.
Because the input fields are isolated samples from the underlying stochastic process, this can easily lead to mismatched features or missed assignments, especially if the number of fields is small or their variability large.
We build on a more comprehensive description of an uncertain scalar field by analyzing the space of \emph{all} its outcomes rather than a finite sample. 

\noindent\textbf{Topology Tracking in Spatio-Temporal Data.}
Several approaches have been described for tracking the temporal evolution of graph-based descriptions of a scalar field's topology over time. 
Szymczak~\cite{Szymczak2005} introduces \emph{subdomain-aware CTs} to analyze the accumulated topological changes to a contour within a user-selected time range. His system allows querying, for instance, for contours that, at the end of the interval, will have split into a certain number of different contours.
Edelsbrunner \etal~\cite{Edelsbrunner2008} use \emph{Jacobi sets} \cite{Edelsbrunner2004} to track a field's Reeb graph over time and describe an algorithm to compute the time-varying Reeb graph of a PL time-dependent scalar field.
Mascarenhas~\cite{Mascarenhas2006} reevaluates the use of the PL interpolation of space-time and proposes to use a piecewise-prismatic interpolation instead to avoid special cases and reduce computation times.

Oesterling \etal~\cite{Oesterling2017} track the evolution of MTs by identifying a sequence of structural changes that must be applied to the MT of one time step to obtain the MT at the next.
Widanagamaachchi \etal~\cite{Widanagamaachchi2012} store the evolution of the field's MT in a \emph{meta graph}, from which \emph{tracking graphs} for an arbitrary isovalues can be extracted. Similarly, Lukasczyk \etal~\cite{Lukasczyk2017} compute a \emph{nested tracking graph}, which describes the temporal evolution of superlevel sets for different isovalues using so-called \emph{nesting trees}. 
Saikia and Weinkauf~\cite{Saikia2017} propose to also consider features extracted from distant time steps to obtain feature tracks that are less sensitive to noise in the data. 

Soler \etal~\cite{Soler2018} use a lifted version of the Wasserstein distance between persistence diagrams~\cite{Turner2014} to find an optimal assignment between the persistence pairs of adjacent time steps. 

Hanser et al. \cite{Hanser2018} investigate the sensitivity of the topology of 1-parametric time-dependent vector fields to parameter changes.

It is worth noting that, while the topology tracking over the realizations of an uncertain field 
and the time steps of a time-dependent field have similarities, methods for one problem are not generally applicable to the other, not least because of profound differences in the way we interpret temporal and stochastic variation.
Still, it may be possible to extend ideas for tracking critical points in time-dependent data to uncertain fields. For example, we were able to adapt the theory and algorithms for temporal tracking based on Jacobi sets by Edelsbrunner \etal~\cite{Edelsbrunner2008} and Mascarenhas~\cite{Mascarenhas2006} to track critical points over the realizations of random fields (see \cref{sec:multiparameter-families,sec:algo-discr}).

%% file: sections/Background.tex
\section{Background}
\label{sec:background}
\noindent\textbf{Critical Points of Deterministic Scalar Fields.}
Let $\dom$ be a differentiable $\domainDims$-manifold and $f: \dom \to \reals$ a twice continuously differentiable \emph{(deterministic) scalar field} on $\dom$. A point $\vec{\gridPt} \in \dom$ is called \emph{critical} if the gradient of $f$ at $\vec{\gridPt}$ vanishes: $\nabla f(\vec{\gridPt}):= \big( \frac{\partial f}{\partial x_i}(\vec{\gridPt})\big)_{i=1}^\domainDims = \vec{0}$; otherwise, it is called \emph{regular}. 
A critical point is \emph{degenerate} if the Hessian matrix of $f$ at $\vec{\gridPt}$, $\mat{H}_f(\vec{\gridPt}):= \big(\frac{\partial^2 f}{\partial x_i\partial x_j} (\vec{\gridPt})\big)_{i,j=1}^\domainDims $, is singular. 
Nondegenerate critical points can be further categorized as maxima, minima, and saddles based on the signs of the eigenvalues of the Hessian matrix. If $f$ has only nondegenerate critical points, it is called \emph{Morse}.

In practice, an explicit description of $f$ is often not available; only the field's values at the vertices $\vec{\gridPt_1},\dots, \vec{\gridPt_\numGridPts} \in \dom$ of some grid $\grid$ are known. We will represent such fields as vectors $\vec{f} \in \reals^\numGridPts$, where the $i^{\text{th}}$ component $f_i$ is the value at grid point $\vec{\gridPt_i}$. 
If $\grid$ consists solely of simplicial cells (e.g., triangles in 2D), we can use piecewise-linear (PL) interpolation to obtain a function defined on the entire domain $\dom$. 
Since PL functions are generally not differentiable, the above definition of critical points is not applicable. An alternative definition for critical points of such fields was given by Banchoff~\cite{Banchoff1970}: For any grid vertex $\vec{\gridPt}$, the collection of its direct neighbors in the grid together with the faces spawned by those vertices is called the \emph{link} of $\vec{\gridPt}$. The portions of the link spawned by neighboring vertices of higher or lower function values are called the \emph{upper} and \emph{lower link}, respectively.
The critical type of $\vec{\gridPt}$ is then determined by the structure of its upper and lower link: If all neighboring vertices have higher/lower function values resulting in an empty lower/upper link, $\vec{\gridPt}$ is a minimum/maximum. If both the lower and the upper link consist of a single maximally connected component, $\vec{\gridPt}$ is regular. Otherwise, $\vec{\gridPt}$ is a saddle point.

\noindent\textbf{Uncertain Scalar Fields and Realization Space.}
In \emph{uncertain} (or \emph{random}) \emph{scalar fields}, the field's value at each location is not known a priori but rather follows an (often unknown) probability density. This means that the field can take on different values, each occurring with a certain probability. Formally, such a field can be thought of as a random variable $F : \Omega \to \reals^{\dom}$ mapping each outcome $\omega$ of a sample space $\Omega$ to a scalar field $f:\dom \to \reals$ over $\dom$, which we call a \emph{realization} of $F$~\cite{Adler2010}. We call the set of all realizations $\realSpace := F(\Omega)$ the \emph{realization space} of $F$. Since we are interested in critical points of such fields, we consider random fields whose realizations are twice continuously differentiable; that is, $\realSpace \subseteq C^2(\dom, \reals)$ (the space of all twice continuously differentiable, real-valued functions on $\dom$).

Again, in practice, a dense description of $F$ at each location is seldom available. As in the deterministic case, such fields are often represented by only describing the values at the $\numGridPts$ vertices of a grid $\grid$ through random variables. To account for spatial autocorrelations, we may think of the random field $F$ as an $\numGridPts$-dimensional multivariate random variable $F: \Omega \to \reals^\numGridPts$ mapping each outcome $\omega$ of $\Omega$ to a discrete field $F(\omega) \in \reals^\numGridPts$. Again the $i^\text{th}$ component of $F(\omega)$ is the value at grid point $\vec{\gridPt_i}$. In this case, the realization space $\realSpace = F(\Omega)$ of $F$ is a subset of the $n$-dimensional Euclidean space $\reals^\numGridPts$.

\noindent\textbf{Jacobi Set.} Let $\fun_i: \domJacobi \to \reals$, $i=0,\dots, \numFields$, be a family of $\numFields+1 \leq \ell$ Morse functions on some differentiable $\ell$-manifold $\domJacobi$. The \emph{Jacobi set} of those functions is defined as the set of points where their gradients are linearly dependent~\cite{Edelsbrunner2004}:
\begin{equation}
\label{eq:def-jacobi-set}
    \jacobi(f_0,\dots, f_\numFields):= \{\vec{\gridPt} \in \domJacobi \mid \rank(\gradMat(\vec{\gridPt})) \leq \numFields \},
\end{equation}
where $\gradMat(\vec{\gridPt}) :=(\nabla f_0(\vec{\gridPt})\ \dots\ \nabla f_\numFields(\vec{\gridPt}))$. 

%% file: sections/Methods.tex
\section{Critical Points with Spatial Uncertainty}
\label{sec:theoretical-considerations}
In uncertain fields, each realization has its own critical points, resulting in a distribution of critical points over the domain. Imagine an uncertain field whose realizations are slightly perturbed versions of a deterministic field. If the perturbation is small compared to the persistence of the base field's critical points, the spatial distribution of the critical points of all realizations will be concentrated around the critical points of the original field (see \cref{fig:shortcomings-local-methods} right). 
In the following, we formalize this intuitive notion of a critical point with spatial uncertainty.

\subsection{Uncertain Critical Points}
\label{sec:uncertain-critical-points}
Given an uncertain scalar field $F: \Omega \to C^2(\dom, \reals)$, a natural starting point for our discussion is the set of all nondegenerate critical points of all its realizations
\begin{equation}
\label{eq:set-of-non-degenerate-crit-pts}
    \ndCritPts := \{(\vec{x}, f) \in \dom \times \realSpace \mid \nabla f(\vec{\gridPt}) = \vec{0}, \det(\mat{H}_f(\vec{\gridPt}))\neq 0\}.
\end{equation}
To identify $F$'s critical points with spatial uncertainty, it is sensible to search for subsets of $\ndCritPts$ consisting of critical points of different realizations that we assume to describe the same topological feature. This raises the question of when two critical points of different realizations can be considered to describe the same feature. Put in mathematical terms, we seek a binary relation $\sameness$ on $\ndCritPts$ that connects two critical points if and only if they are manifestations of the same topological feature. The expression ``manifestation of the \emph{same} topological feature'' immediately suggests that $\sameness$ should be an equivalence relation (i.e., symmetric, transitive, and reflexive).
The equivalence classes of such a relation then yield a partition of $\ndCritPts$ into subsets of critical points that are considered manifestations of the same topological feature.  Because, intuitively, a nondegenerate critical point occurring at $\vec{\gridPt}$ in some realization $f$ of $F$ is a manifestation of exactly one feature, we call the equivalence class $[(\vec{\gridPt}, f)]_\sameness \subseteq \ndCritPts$, consisting of all critical points that are manifestations of the same feature, an \emph{uncertain critical point} of $F$. Accordingly, we call each element of such an equivalence class a \emph{manifestation} of that uncertain critical point.

It is not immediately clear what a sensible definition of $\sameness$ should look like. 
The motivation for a first idea has already been addressed briefly in the introduction: Due to the random nature of the investigated phenomena, even slightest perturbations of the initial conditions will cause a ``likely critical point'' to occur at marginally different locations in different realizations (see \cref{fig:shortcomings-local-methods} right).
This suggests considering critical points to be manifestations of the same uncertain critical point if they are part of the same connected component in $\dom$. This concept of spatial closeness has already been used in clustering approaches to identify common critical points in ensembles of fields~\cite{Favelier2019,Kappe2022}.

However, treating critical points of different realizations as manifestations of the same uncertain critical point solely based on their spatial connectedness can produce misleading results as it might connect unrelated features.
To make this more clear, imagine a time-dependent scalar field where two equivalent features revolve with constant radius and speed around their fixed centroid. The collection of possible locations of both features in the domain is a single circle from which the two features can no longer be distinguished. To overcome this issue, methods for temporal tracking make use of the temporal coherency between time steps. That is, under the basic assumption that a small change in time will result in a small movement of the feature, they only consider two features the same if one can be continuously transformed into the other when proceeding continuously through time.

To apply this idea to uncertain fields, we make the following intuitive continuity assumption:
For any realization in $\realSpace$ there is another realization with an arbitrarily small distance (i.e., differing only infinitesimally). That is, we assume that the realizations of $F$ are not isolated points but form dense regions in $C^2(\dom, \reals)$. Note that this requirement is not specific to our work but is common in uncertainty visualization (e.g., \cite{Kao2002,Poethkow2010,Schlegel2012,Poethkow2013}).
From this assumption, it follows that for each critical point $\vec{\gridPt}$ of a realization $f$, we find other realizations with an arbitrarily small distance to $f$, which, because they differ only slightly from $f$, also have a critical point located arbitrarily close to $\vec{\gridPt}$.
In other words, the locations of critical points will vary continuously when continuously moving through realization space.
The elements of $\ndCritPts$ are then also not isolated points but form dense regions in $\dom \times \realSpace$.
In analogy to the temporal tracking, it therefore makes sense to declare two critical points in two realizations manifestations of the same uncertain critical point if they are connected by a series of manifestations within $\ndCritPts$ itself.

As in temporal tracking, there generally will be realizations in $\realSpace$ where the number of features of the random field changes. Consider, for instance, the birth of an extremum saddle pair in some realization $f$ at location $\vec{\gridPt}$. Moving away from realization $f$ the newly created extremum and saddle will move in $\dom$ resulting in paths describing their evolution in $\ndCritPts$ connected by the point of generation $(f, \vec{\gridPt}) \in \ndCritPts$. That is, the extremum and the saddle are connected by a series of manifestations in $\ndCritPts$. However, because they have different types, they obviously should not be considered manifestations of the same feature. Considering this, we propose the following definition of $\sameness$:

\begin{definition}
\label{def:sameness-relation}
Let $F: \Omega \to C^2(\dom, \reals)$ be an uncertain scalar field with twice continuously differentiable realizations, and define $\ndCritPts$ according to \cref{eq:set-of-non-degenerate-crit-pts}. 
 Critical points $\vec{\gridPt_1}$ and $\vec{\gridPt_2}$ of two realizations $f_1$ and $f_2$, respectively, are considered \emph{manifestations of the same feature} (i.e., $(\vec{\gridPt_1}, f_1) \sameness (\vec{\gridPt_2}, f_2)$) if the critical point at $\vec{\gridPt_1}$ can be continuously traced to the one at $\vec{\gridPt_2}$ while continuously altering the realization from $f_1$ to $f_2$ and maintaining the type of critical point---that is, if there is a continuous path between $(\vec{\gridPt_1}, f_1)$ and $(\vec{\gridPt_2}, f_2)$ with constant critical point type within $\ndCritPts$.
\end{definition}
The resulting equivalence classes of $\sameness$ then yield our definition of uncertain critical points: 
\begin{definition}
\label{def:uncertain-critical-point}
In the setting of \cref{def:sameness-relation}, we call each maximally connected component of $\ndCritPts$ with constant critical point type---constituting one equivalence class $[(\vec{\gridPt}^*, f^*)]_\sameness$ of some representative element $(\vec{\gridPt}^*, f^*) \in \ndCritPts$ with respect to $\sameness$---an \emph{uncertain critical point} of $F$.
\end{definition}

\subsection{Spatial Distribution}
\label{sec:spatial-distribution}
To achieve our initial goal of identifying critical points with spatial uncertainty, we can project each uncertain critical point---living in $\ndCritPts \subseteq \dom \times \realSpace$---to the domain. Because, by our previous construction, each uncertain critical point is connected in $\ndCritPts$, its projection will also be a connected region within $\dom$. Furthermore, it would be desirable if one could specify the probability distribution of an uncertain critical point over this region. Using the fact that $F$ is a random variable mapping each point of a sample space $\Omega$ to a realization $f\in\realSpace$, we can find the probability for an uncertain critical point to manifest in some part of its spatial projection.
Let $\ucp \subseteq \ndCritPts$ be an uncertain critical point of $F$. The probability of observing a manifestation of $\ucp$ in some region $D \subseteq \dom$ is then obtained as the probability of observing a realization $f\in \realSpace$ that has a critical point at $\vec{\gridPt} \in D$ that is a manifestation of $\ucp$ (i.e., $(\vec{\gridPt}, f) \in \ucp$):
\begin{equation}%
 \prob(\ucp \text{ in } D) = \prob_\Omega(F^{-1}(\{f \in \realSpace\mid \exists \vec{\gridPt} \in D: (\vec{\gridPt}, f) \in \ucp\})),
 \label{eq:probability-of-region}
\end{equation}%
where $\prob_\Omega$ is a suitable probability measure on $\Omega$.

It is worth noting that the two events ``$\ucp$ in $D_1$'' and ``$\ucp$ in $D_2$'' are not necessarily mutually exclusive, even when considering two disjoint regions $D_1$ and $D_2$. In particular, in the inclusion-exclusion principle, $\prob(\ucp\text{ in } D_1 \cup D_2) = \prob(\ucp \text{ in }D_1) + \prob(\ucp \text{ in } D_2) - \prob(\ucp \text{ in } D_1 \land \ucp \text{ in } D_2)$, the last term might be non-zero.
Note that in this case $\prob(\ucp \text{ in } \cdot)$ is not $\sigma$-additive and hence does not characterize a probability distribution.
Looking at \cref{eq:probability-of-region}, it seems especially likely that $\prob(\ucp \text{ in }D_1 \land \ucp \text{ in } D_2) \neq 0$ if already the sets of corresponding realizations $\{f \in \realSpace\mid \exists \vec{\gridPt} \in D_i: (\vec{\gridPt}, f) \in \ucp\}$, $i=1,2$, of the two events have a non-negligible intersection---that is, if there is a considerable number of realizations having a manifestation of $\ucp$ in $D_1$ as well as $D_2$. An explicit example of this is given in the next section (\cref{example:counterexample}), which will also highlight entailed challenges for the characterization of the spatial uncertainty of a critical point.

\section{Special Case: Multiparameter Families of Fields}
\label{sec:multiparameter-families}
In this section, we concentrate on random fields whose realization space can be parameterized by a finite set of parameters; that is, $F$ now maps each element $\omega$ of the sample space $\Omega$ to a realization $f_\vec{\param}: \dom \to \reals$, which is uniquely described by a parameter vector $\vec{\param}$ stemming from an $\numFields$-dimensional \emph{parameter space} $\paramSpace := \reals^\numFields$. We will further assume that $f(\vec{\gridPt}, \vec{\param}):=f_\vec{\param}(\vec{\gridPt})$ is sufficiently smooth in both $\vec{\gridPt}$ and $\vec{\param}$, thus also ensuring the continuity assumption made in the previous section. The random nature of $F$ can then be fully described by taking the parameter vectors $\vec{\param}$ as the outcomes of a multivariate random variable $\vec{\rvParam}: \Omega \to \paramSpace$. 

While at first glance the restriction to this class seems to be a significant limitation, it still contains several random fields with practical relevance~\cite{Adler} including random polynomials~\cite{BharuchaReid2014}, the cosine process/field~\cite[Sec. 14.4.4]{Adler2010}, and orthogonal expansions of centered Gaussian processes~\cite[Chapt. 3]{Adler2010}.
Finally, and probably most relevant for the field of visualization, any field obtained from interpolating between values at $\numGridPts$ discrete grid points can be described by a weighted sum of basis functions, weighted by the respective value at each grid point. Taking those values as the outcomes $\vec{\param}$ of an $\numGridPts$-dimensional random variable $\vec{\rvParam}$ then also yields a parametric random field $f_\vec{\param}: \dom \to \reals$.

\begin{figure}
    \centering
   \begin{tikzpicture}
        \node[inner sep=0pt] (jacobi-3d) {\hspace*{0em}\includegraphics[width=0.43\linewidth]{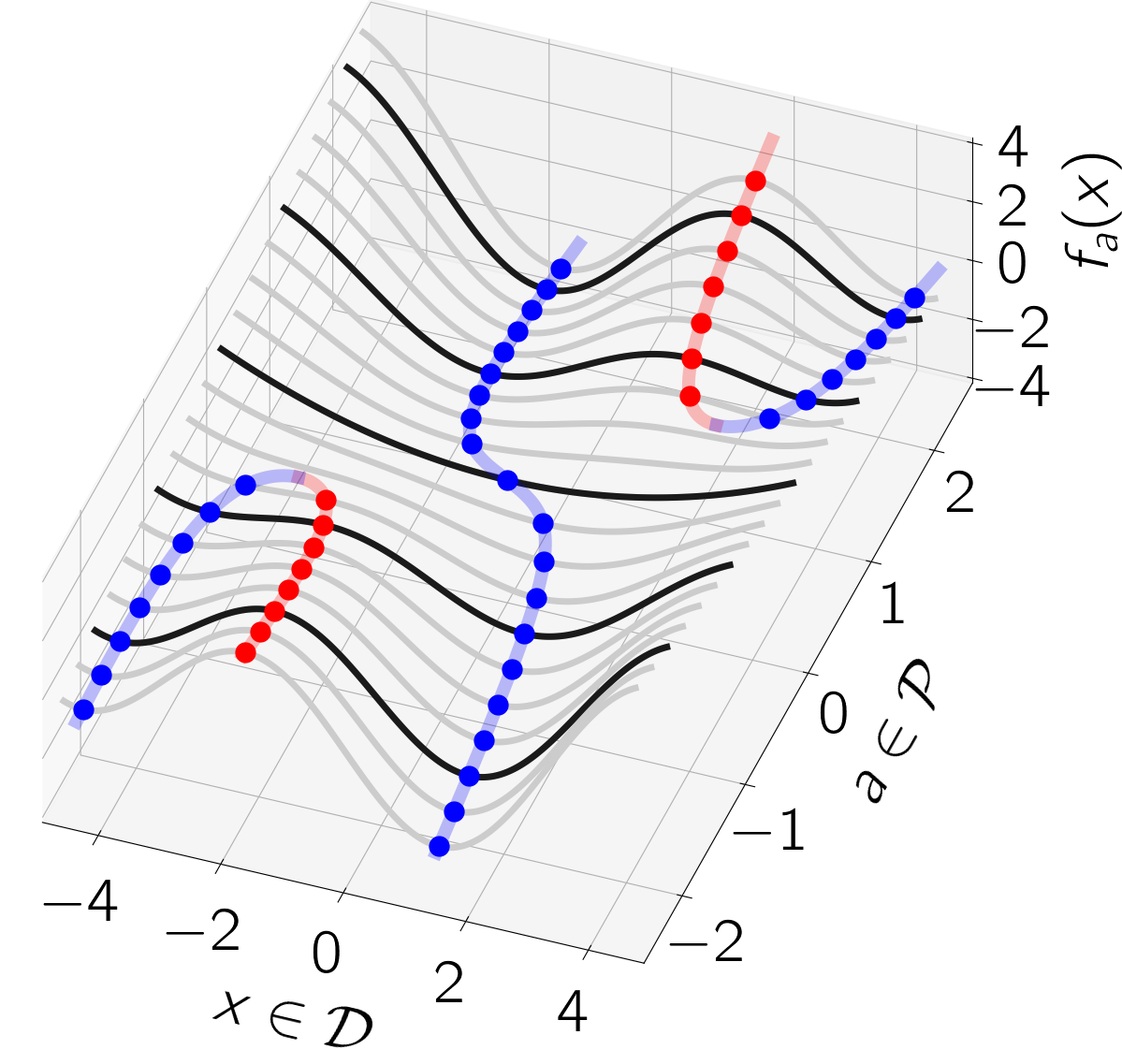}};
        \node[inner sep=0pt, right=1em of jacobi-3d.north east, anchor=north west] (jacobi-2d) {\includegraphics[width=0.5\linewidth]{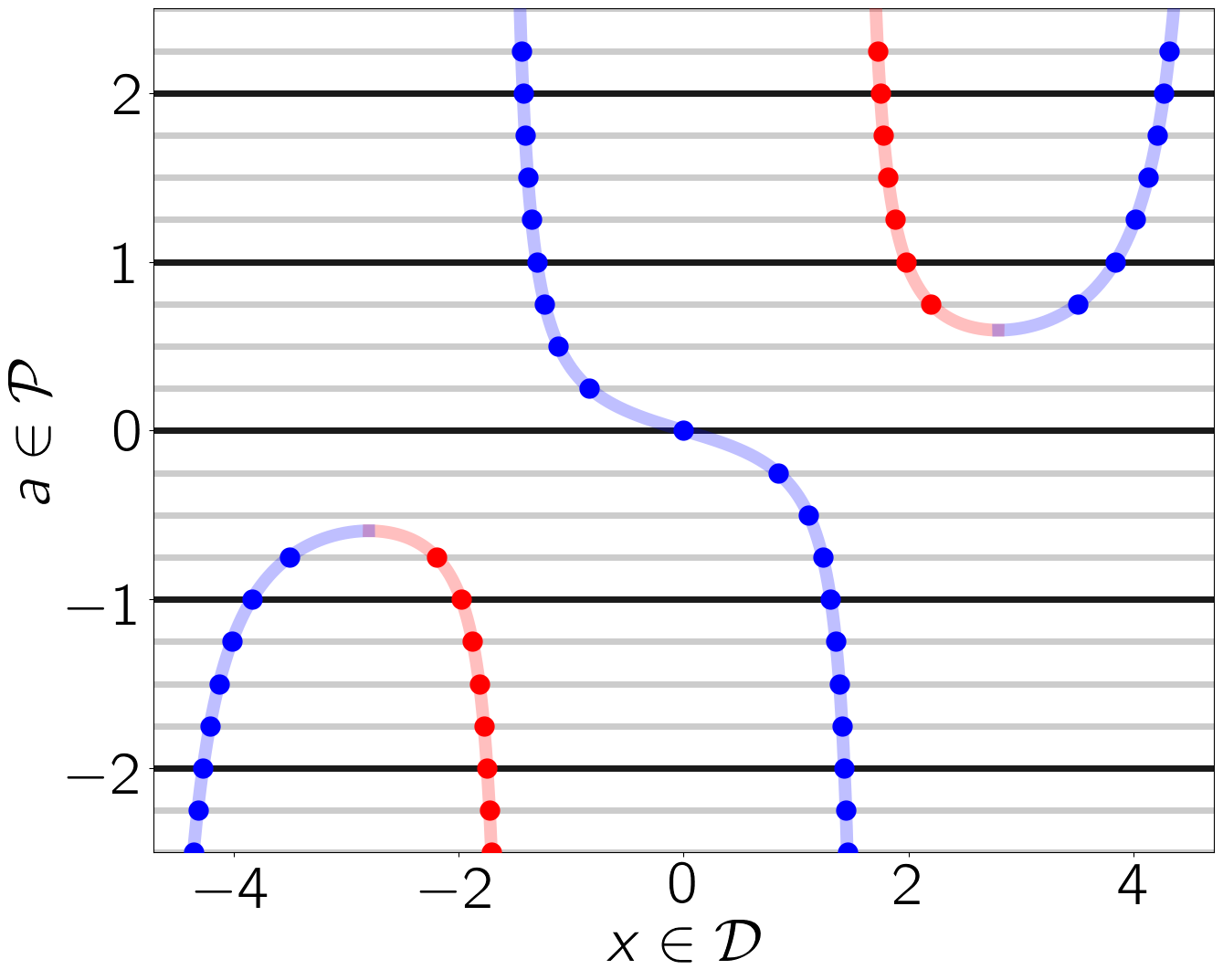}};
        
        \node[left=0em of jacobi-3d.north west, anchor=north west] {(a)};
        \node[left=1em of jacobi-2d.north west, anchor=north west] {(b)};
    \end{tikzpicture}
      \caption{(a) Realizations of the one-parameter family of functions $f_\param(\gridPt) = \frac{\gridPt^2}{10} + \param \sin(\gridPt)$ with their maxima and minima highlighted in red and blue, respectively. (b) Projection of the critical points onto the ambient space $\dom \times \paramSpace$. Semi-transparent lines visualize the extrema/critical points of the intermediate realization.}
    \label{fig:1D-1D-jacobi}
\end{figure}

For a first intuition, consider the simple example of the one-dimensional, one-parametric function depicted in \cref{fig:1D-1D-jacobi}. The 3D plot of $f_\param(\gridPt)$ on the left illustrates the continuous evolution of the realizations (gray and black curves) for different parameter values. The maxima and minima of each realization are marked in red and blue, respectively. 
By projecting the found extrema of all realizations onto the space spanned by the domain $\dom$ and the parameter space $\paramSpace$ (we call this the \emph{ambient space}), we obtain \crefSubFigRef{fig:1D-1D-jacobi}{b}. Note that the projected points make up precisely the portion of $\ndCritPts$ within the depicted area (see \cref{eq:set-of-non-degenerate-crit-pts}).
In this plot, we can make out five maximally connected components with constant critical point types: three comprising minima and two comprising maxima. Those are the five uncertain critical points of $f_\param$ in the depicted region.

\subsection{Relation to Jacobi Sets}
\label{sec:relation-to-jacobi-sets}
The above example suggests that uncertain critical points are smooth, connected regions embedded in the ambient space.
We now prove this observation for the general case of multiparameter families of fields $f_\vec{\param}$. For this, we show how the concept of uncertain critical points relates to the Jacobi set of $f_\vec{\param}$ and a set of auxiliary functions. Let
\begin{equation}
    \label{eq:set-of-crit-pts}
    \jacobi_{\fun_\vec{\param}}:= \{(\vec{\gridPt}, \vec{\param}) \in \dom \times \paramSpace \mid \nabla f_\vec{\param}(\vec{\gridPt}) = \vec{0}\}
\end{equation}
be the set of \emph{all} critical points (including nondegenerate ones, unlike $\ndCritPts$) in all realizations of $\fun_\vec{\param}$ (see \crefSubFigRef{fig:1D-1D-jacobi}{b}).
For the case $m=1$ (i.e., a one-parameter family of functions), Edelsbrunner \etal~\cite{Edelsbrunner2004, Edelsbrunner2008} showed that this set is related to the Jacobi set of a collection of Morse functions.
We extend this observation to multiparameter families of functions:
\begin{lemma}
\label{lemma:relation-to-jacobi-set}
Let $\helper_0(\vec{\gridPt}, \vec{\param}):= \fun_\vec{\param}(\vec{\gridPt})$ be a generic Morse function on $\dom \times \paramSpace$ and define the auxiliary functions $\helper_i(\vec{\gridPt}, \vec{\param}):= \param_i, i=1,\dots, \numFields$. Then $\jacobi_{\fun_\vec{\param}}$ is the Jacobi set of $\helper_0,\dots, \helper_\numFields$.
\end{lemma}
\begin{Proof}
For $i = 1, \ldots, \numFields$, we have $\nabla \helper_i(\vec{\gridPt}, \vec{\param}) = \vec{e_{\domainDims+i}}$, the $(\domainDims+i)^\text{th}$ canonical basis vector. The definition of the Jacobi set (\cref{eq:def-jacobi-set}) yields $\jacobi(\helper_0, \dots, \helper_\numFields) = \{(\vec{\gridPt}, \vec{\param}) \in \dom\times \paramSpace\mid \rank(\gradMat(\vec{\gridPt}, \vec{\param})) \leq \numFields\}$ with $\gradMat(\vec{\gridPt}, \vec{\param}) = (\nabla \helper_0(\vec{\gridPt}, \vec{\param})\ \vec{e_{\domainDims+1}}\ \dots\ \vec{e_{\domainDims+\numFields}})$. Since the last $\numFields$ columns of this matrix are linearly independent, it has rank at least $\numFields$. Moreover, it has rank larger than $\numFields$ if and only if $\nabla_\vec{\gridPt} \helper_0(\vec{\gridPt}, \vec{\param}) = \nabla \fun_{\vec{\param}}(\vec{\gridPt}) \neq \vec{0}$. Thus $\jacobi(\helper_0,\dots, \helper_\numFields) = \{(\vec{\gridPt}, \vec{\param}) \in \dom\times \paramSpace\mid \nabla \fun_\vec{\param}(\vec{\gridPt}) = \vec{0}\} = \jacobi_{f_\vec{\param}}$.
\end{Proof}
We can use that connection to show that $\jacobi_{f_\vec{\param}}$ has a suitable structure that facilitates the identification of uncertain critical points as smooth connected regions as suggested by the example above:
\begin{proposition}
\label{proposition:smooth-submanifold}
If $\helper_0(\vec{\gridPt}, \vec{\param}):= \fun_\vec{\param}(\vec{\gridPt})$ is a generic Morse function on $\dom \times \paramSpace$, then $\jacobi_{\fun_\vec{\param}}$ is a smooth $\numFields$-dimensional submanifold of $\dom \times \paramSpace$.
\end{proposition}
\begin{Proof}
According to~\cref{lemma:relation-to-jacobi-set}, $\jacobi_{\fun_{\vec{\param}}}$ is the Jacobi set of the functions $\helper_0,\ldots,\helper_\numFields$. Edelsbrunner and Harer~\cite{Edelsbrunner2004} gave a sufficient condition for the Jacobi set of $\numFields+1$ Morse functions to be a smoothly embedded $\numFields$-dimensional submanifold of their domain. Applied to our case, $\jacobi_{\fun_\vec{\param}}=\jacobi(\helper_0,\dots, \helper_\numFields)$ is a smooth $m$-dimensional submanifold of $\dom \times \paramSpace$ if the set $S_2:= \{(\vec{\gridPt}, \vec{\param}) \in \dom \times \paramSpace\mid \rank(\gradMat(\vec{\gridPt}, \vec{\param})) \leq m-1\}$ with $\gradMat(\vec{\gridPt}, \vec{\param}) = (\nabla \helper_0(\vec{\gridPt}, \vec{\param})\ \vec{e_{\domainDims+1}}\ \dots\ \vec{e_{\domainDims+\numFields}})$ is empty.  We already saw in the proof of~\cref{lemma:relation-to-jacobi-set} that $\gradMat(\vec{\gridPt}, \vec{\param})$ has rank at least $\numFields$ for all $(\vec{\gridPt}, \vec{\param}) \in \dom\times \paramSpace$. That is, $S_2$ is indeed empty.
\end{Proof}
\begin{remark}
The term ``generic'' in \cref{lemma:relation-to-jacobi-set} and \cref{proposition:smooth-submanifold} refers to the fact that there are Morse functions for which the Jacobi set is not guaranteed to be an $\numFields$-manifold everywhere. Often, however, this is only the case at a negligible number of discrete points and can be resolved by a slight perturbation of the function. An example of this is discussed in the supplemental material.
\end{remark}

In the basic example from \cref{fig:1D-1D-jacobi}, $f_\param$ is a linear one-parameter function family---that is, of the form $\fun_\param(\gridPt) = \evs_0(\gridPt) + \param \evs_1(\gridPt)$. The condition $\fun_\param'(\gridPt) = 0$ in the definition of $\jacobi_{\fun_\param}$ (\cref{eq:set-of-crit-pts}) then yields $\evs_0'(\gridPt) = \evs_1'(\gridPt) = 0$ or $\param = -\evs_0'(\gridPt)/\evs_1'(\gridPt)$. That is, aside from points where the derivatives of $\evs_0$ and $\evs_1$ vanish simultaneously, $\jacobi_{\fun_\param}$ is described explicitly by the graph of the function $\param(\gridPt) = -\evs_0'(\gridPt)/\evs_1'(\gridPt)$.
\CrefSubFigRef{fig:1D-1D}{a} depicts this graph for the example from \cref{fig:1D-1D-jacobi} (i.e.,  for $\evs_0(\gridPt) = \frac{\gridPt^2}{10}$ and $\evs_1(\gridPt) = \sin(\gridPt)$).
There, one sees that poles (vertical black lines) occur in $\jacobi_{\fun_\param}$ at the roots of $\evs_1'(\gridPt) = \cos(\gridPt)$. These poles partition $\jacobi_{\fun_\param}$ into multiple connected components, tracing different critical points across the realizations of $\fun_\param$. Moreover, by coloring each point of $\jacobi_{\fun_\param}$ in red or blue (ignoring the specific color mapping for the moment) if $\fun_\param(\gridPt)$ attains a maximum or minimum at that point, we find that those components can be further partitioned into branches of constant critical type.
In the general setting, we have:
\begin{proposition}
\label{proposition:exists-point-where-det-hessian-zero}
    Let $\helper_0(\vec{\gridPt}, \vec{\param}) := \fun_\vec{\param}(\gridPt)$ be a generic Morse function that is twice continuously differentiable w.r.t. $\vec{\gridPt}$ with  derivatives that are continuous in $\vec{\param}$ and let $(\vec{\gridPt_1}, \vec{\param_1}), (\vec{\gridPt_2}, \vec{\param_2})\in \dom \times\paramSpace$ be two points in the same connected component of $\jacobi_{\fun_\vec{\param}}$ such that $\fun_\vec{\param_1}$ has a different type of critical point at $\vec{\gridPt_1}$ than $\fun_\vec{\param_2}$ at $\vec{\gridPt_2}$. Then, on each continuous path connecting those points within that component, there is a point $(\vec{\gridPt^*}, \vec{\param^*})$ such that $\det(\mat{H}_{\fun_{\vec{\param^*}}}(\vec{\gridPt^*})) = 0$.
\end{proposition}
\begin{Proof}
    Let $\mathcal{C}\subset \jacobi_{\fun_\vec{\param}}$ be the connected component containing $(\vec{\gridPt_1}, \vec{\param_1}), (\vec{\gridPt_2}, \vec{\param_2})$, and $\gamma : [0,1] \to \mathcal{C}, \gamma(t) \mapsto (\vec{\gridPt}(t), \vec{\param}(t))$ an arbitrary continuous path connecting them; that is, $\gamma(0) = (\vec{\gridPt_1}, \vec{\param_1})$ and $\gamma(1) = (\vec{\gridPt_2}, \vec{\param_2})$. Because $\fun_\vec{\param}$'s second-order derivatives are continuous in $\vec{\gridPt}$ and $\vec{\param}$ and $\gamma$ is continuous, $\mat{H}_{\fun_{\vec{\param}(t)}}(\vec{\gridPt}(t))$ is a continuous function from $[0,1]$ to $\reals^{\domainDims\times \domainDims}$. We thus have $\domainDims$ continuous functions $\lambda_i: [0,1]\to \reals$ such that for each $t \in [0,1]$ $\{\lambda_1(t), \dots, \lambda_\domainDims(t)\}$ are the eigenvalues of $\mat{H}_{\fun_{\vec{\param}(t)}}(\vec{\gridPt}(t))$ (see, e.g.,\cite[Theorem~5.2]{Kato2012}). The fact that $\fun_{\vec{\param_1}}$ has a different type of critical point at $\vec{\gridPt_1}$ than $\fun_{\vec{\param_2}}$ at $\vec{\gridPt_2}$ yields that $\mat{H}_{\fun_{\vec{\param_1}}}(\vec{\gridPt_1})$ and $\mat{H}_{\fun_{\vec{\param_2}}}(\vec{\gridPt_2})$ have at least one eigenvalue $\lambda_j$ with different sign. 
    Then, according to the intermediate value theorem, there is $t^* \in (0,1)$ where $\lambda_j(t^*) = 0$. Thus, at $(\vec{\gridPt^*}, \vec{\param^*}) := \gamma(t^*)$, we have $\det(\mat{H}_{\fun_\vec{\param^*}}(\vec{\gridPt^*})) = \prod_{i=1}^{\domainDims} \lambda_i(t^*) = 0$.
\end{Proof}

\pgfdeclarelayer{bg} 
\pgfsetlayers{bg,main} 

\begin{figure}[t]%
  \centering
    \begin{tikzpicture}
        \pdfpxdimen=\dimexpr 1in/300\relax
        \node[black, thick, inner sep=0pt] (continuous) {%
            \includegraphics[%
                width=\linewidth%
            ]{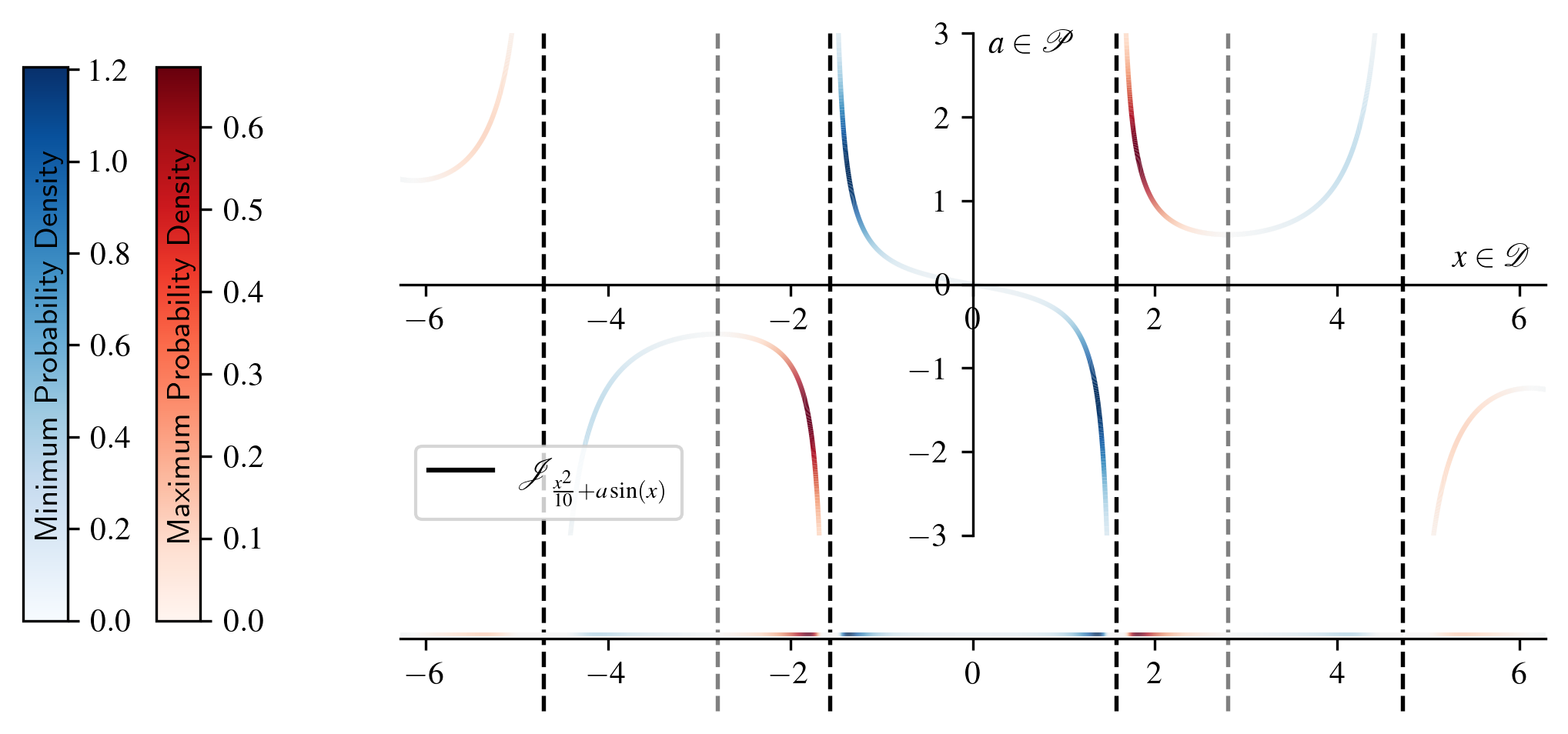}};
            
        \coordinate (roi-south-west) at ($(continuous.south west) + (19.2em,5em)$);
        \coordinate (roi-north-east) at ($(continuous.north east) - (6.5em,2.2em)$);
        
        \draw[black, rounded corners=2pt] ($(continuous.south west) + (5em, 3em)$) rectangle (continuous.north east);
        \draw[black, rounded corners=2pt] ($(continuous.south west) + (5em, 0.1em)$) rectangle ($(continuous.south east) + (0em, 2.3em)$);
   
        \node[below right=.1em and 4.9em of continuous.north west, anchor=north west] (labelA) {(a)};
        \node[above right=.7em and 4.9em of continuous.south west, anchor=south west] (labelB) {(b)};
    \end{tikzpicture}
  \caption{(a) Set of critical points of $\fun_\param(\gridPt):= \frac{x^2}{10}+\param\sin(\gridPt)$. Vertical black lines mark poles that decompose $\jacobi_{\fun_{\param}}$ into connected components. Vertical gray lines mark domain points where the Hessian is singular, indicating points where the critical point type changes. The color map encodes the probability density within each branch of constant type, indicating how likely it is that the critical point traced by this branch is located at any position. (b) Probability-weighted projection of $\jacobi_{\fun_\param}$ onto $\dom$ reveals the spatial distribution of each uncertain critical point.}
  \label{fig:1D-1D}
\end{figure}
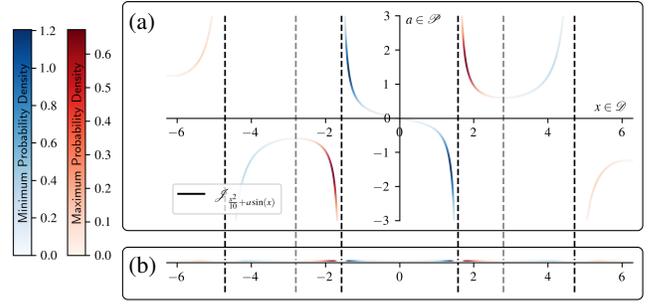%

\Cref{proposition:exists-point-where-det-hessian-zero} tells us that we can indeed further subdivide the connected components of $\jacobi_{\fun_\vec{\param}}$ at all points where the Hessian of $\fun_\vec{\param}$ is singular 
to obtain connected components of constant critical point type. Removing those points from $\jacobi_{\fun_\vec{\param}}$ produces exactly the set of all nondegenerate critical points $\ndCritPts$ (see \cref{eq:set-of-non-degenerate-crit-pts}) analyzed in the general setting in \cref{sec:theoretical-considerations}. The resulting subcomponents of constant type of $\jacobi_{\fun_\vec{\param}}$ are hence the uncertain critical points of $\fun_\vec{\param}$ (see \cref{def:uncertain-critical-point}).

\begin{remark}
Partitioning the set of critical points depicted in \crefSubFigRef{fig:1D-1D}{a} at all poles (black lines) and the points where the Hessian is singular (gray lines) results in branches, each representing an uncertain critical point with a certain type. Moreover, those lines partition the domain in disjoint regions, meaning no two uncertain critical points will occur at the same location. This is a consequence of $\jacobi_{\fun_\param}$ having an explicit description as a function over the domain in this linear one-parametric case but is not generally the case for arbitrary multiparameter families of functions. We provide further discussion of this matter in the supplemental material.
\end{remark}

\subsection{Spatial Distribution}
\label{sec:spatial-distribution2}
The parameterized representation of $F$ also allows rephrasing the abstract description of the probability of observing an uncertain critical point $\ucp$ in some spatial region $D\subset \dom$ in \cref{eq:probability-of-region} in an explicit form.
Recall that the parameter $\vec{\param}$ is also the value of a multivariate random variable $\vec{\rvParam}$.
To obtain the probability of $\ucp$ to occur in $D$, we integrate $\vec{\rvParam}$'s PDF $\pdf_\vec{\rvParam}$ over the set of parameters whose corresponding realizations have a critical point within that region that is a manifestation of $\ucp$, $P_\ucp(D):= \{\vec{\param} \in \paramSpace\mid \exists \vec{\gridPt} \in D: (\vec{\gridPt}, \vec{\param})\in \ucp\}$:
\begin{equation}
\label{eq:probability-of-region-parametric}
    \prob(\ucp \text{ in } D) = \int_{P_\ucp(D)} \pdf_\vec{\rvParam}\, \mathrm{d}^\numFields\vec{\param}.
\end{equation}

We again consider the example of a one-dimensional linear one-parametric random field (i.e., $\fun_\param(\gridPt) = \evs_0(\gridPt) + \param \evs_1(\gridPt)$).
Recall that in this case the points on $\jacobi_{\fun_\param}$ generically can be expressed explicitly by $\param(\gridPt) = -\evs_0'(\gridPt)/\evs_1'(\gridPt)$. It is easy to show that the derivative $\param'(\gridPt)$ of this function is zero if and only if the Hessian of the realization $\fun_{\param(\gridPt)}$ is singular at $\gridPt$. Since those points, by construction, are not part of any uncertain critical point of $\fun_\param$, $\param(\gridPt)$ is monotonic on all of them. 
The probability of an uncertain critical point $\ucp$ to manifest itself in an interval $[\gridPt_0, \gridPt_0+h] \subset \{\gridPt \in \dom\mid \exists \param\in \paramSpace: (\gridPt, \param) \in \ucp\}$ is thus given by 
$\prob(\ucp\text{ in } [\gridPt_0, \gridPt_0+h]) = \left| \int_{\param(\gridPt_0)}^{\param(\gridPt_0+h)} \pdf_\rvParam(y)\, \mathrm{d}y \right|
     = \left|\cdf_\rvParam(\param(\gridPt_0+h))-\cdf_\rvParam(\param(\gridPt_0)) \right|$, where $\cdf_\rvParam$ is the cumulative distribution function (CDF) of $\rvParam$. From this, we can compute the spatial probability density of $\ucp$ at $\gridPt_0$ by normalizing this probability with the interval length $h$ and taking the limit $h \to 0$:
\begin{align}
    \label{eq:spatial-density}
    \begin{split}
         \left|\lim_{h\to 0}\right. &\left.\frac{\cdf_\rvParam(\param(\gridPt_0+h))-\cdf_\rvParam(\param(\gridPt_0))}{h}\right|= \left|\frac{\mathrm{d}}{\mathrm{d}\gridPt} \cdf_\rvParam(\param(\gridPt))\big|_{\gridPt=\gridPt_0}\right| \\
         &= |\pdf_\rvParam(\param(\gridPt_0)) \cdot \param'(\gridPt_0)|.
    \end{split}
\end{align}

For the example in \cref{fig:1D-1D}, we exemplarily assumed $\rvParam$ to follow a standard normal distribution and used white-red and white-blue diverging color maps to encode the spatial density given by \cref{eq:spatial-density} along the uncertain maxima and minima, respectively. Projecting $\jacobi_{\fun_\param}$ onto $\dom$ (i.e., the $x$-axis) then yields the density distribution of $\fun_\param$'s critical points with spatial uncertainty (\crefSubFigRef{fig:1D-1D}{b}). A noteworthy property of the found density is that its maximal values do not coincide with the maximum of the density of $\rvParam$. For instance, the uncertain minimum in the region $\gridPt \in \left(-\frac{\pi}{2},\frac{\pi}{2}\right)$ is more likely to occur near the border than the center, where the standard normal distribution attains its maximum at $\param = 0$. 
This demonstrates that using our definition of uncertain critical points, we can also identify critical points with a multimodal spatial distribution.

In the above derivation of the spatial density,
we required the injectivity of $\param(\gridPt)$. This especially ensures that for two disjoint regions $D_1, D_2\subset \dom$, also the sets of realizations where $\ucp$ has a manifestation in one of them, $P_\ucp(D_1)$ and $P_\ucp(D_2)$, are disjoint, which we already identified as a potential necessity for $\prob(\ucp \text{ in } \cdot)$ to characterize a probability distribution over $\dom$ in the general discussion in \cref{sec:spatial-distribution}. In general, we have:

\begin{proposition}
\label{proposition:projectability-to-param-space}
  Let $\helper_0(\vec{\gridPt}, \param):= \fun_\param(\vec{\gridPt})$ be a generic \emph{one-parametric} Morse function. Then each uncertain critical point $\ucp$ in $\jacobi_{\fun_\param}$ has an overlap-free projection onto $\paramSpace$; that is, if $(\vec{\gridPt_1}, \param_1), (\vec{\gridPt_2}, \param_2) \in \ucp$ with $\vec{\gridPt_1} \neq \vec{\gridPt_2}$, then also $\param_1 \neq \param_2$.
\end{proposition}
\begin{Proof}
    Let $(\vec{\gridPt_1}, \param_1), (\vec{\gridPt_2}, \param_2) \in \ucp$ with $\vec{\gridPt_1} \neq \vec{\gridPt_2}$. In the previous subsection, we found that each uncertain critical point is a connected component of constant critical point type of $\jacobi_{\fun_\param}$. The points $(\vec{\gridPt_1}, \param_1)$ and $(\vec{\gridPt_2}, \param_2)$ are, therefore, connected by a path in $\jacobi_{\fun_\param}$ along which $\mat{H}_{\fun_{\param}}$ is regular. Let $\hat{\gamma}: [0,1] \to \ucp, \hat{\gamma}(t) := (\hat{\vec{\gridPt}}(t), \hat{\param}(t))$ be any such path.
    Because $\jacobi_{\fun_\param}$ is a differentiable 1-manifold (\cref{proposition:smooth-submanifold}), $\hat{\gamma}$ is differentiable and there is a regularly reparameterized version $\gamma(t) := (\vec{\gridPt}(t), \param(t))$ of $\hat{\gamma}$ (i.e., $\dot{\gamma}(t)  = (\dot{\vec{\gridPt}}(t), \dot{\param}(t))\neq \vec{0}$ for all $t \in [0,1]$).
    
    Now assume, for a proof by contradiction, that $\param(0) = \param_1 =\param_2 =\param(1)$. Because $\param(t)$ is continuous on $[0,1]$ and differentiable on $(0,1)$, the mean value theorem guarantees the existence of $t^* \in (0,1)$ such that $\dot{a}(t^*) = (a(1) - a(0))/(1-0) = 0$.
    Since $\gamma([0,1]) \subset \jacobi_{\fun_\param}$, we have $\nabla \fun_{\param(t)}(\vec{\gridPt}(t)) = \vec{0}$ (i.e., constant) for all $t\in [0,1]$. Hence,
    $\vec{0} = \frac{\mathrm{d}}{\mathrm{d}t} \nabla_{\vec{\gridPt}} \helper_0(\vec{\gridPt}(t), \param(t))
        = \mat{H}_{\fun_{\param(t)}}(\vec{\gridPt}(t))\cdot \dot{\vec{\gridPt}}(t) + \frac{\partial}{\partial a}\nabla_{\vec{\gridPt}} \helper_0(\vec{\gridPt}(t), \param(t))\cdot \dot{\param}(t)$
    and at $t = t^*$, therefore, $\mat{H}_{\fun_{\param(t^*)}}(\vec{\gridPt}(t^*))\cdot \dot{\vec{\gridPt}}(t^*) = \vec{0}$. Because of the regularity of $\gamma$ and $\dot{\param}(t^*) = 0$, we have $\dot{\vec{\gridPt}}(t^*) \neq \vec{0}$ and hence $\mat{H}_{\fun_{\param(t^*)}}(\vec{\gridPt}(t^*))$ is singular; that is, there is a point along $\gamma$ and therewith along $\hat{\gamma}$ where $\mat{H}_{\fun_{\param}}$ is singular. Because $\hat{\gamma}$ was arbitrary, $(\vec{\gridPt_1}, \param_1)$ and $(\vec{\gridPt_2}, \param_2)$ cannot be part of the same uncertain critical point. A contradiction, and hence $\param_1 = \param(0) \neq \param(1) = \param_2$.
\end{Proof}

\begin{figure}
  \centering
    \begin{tikzpicture}
        \pdfpxdimen=\dimexpr 1in/300\relax
        \node[black, thick, inner sep=0pt] (3d-view) {%
            \includegraphics[%
                height=4.3cm
            ]{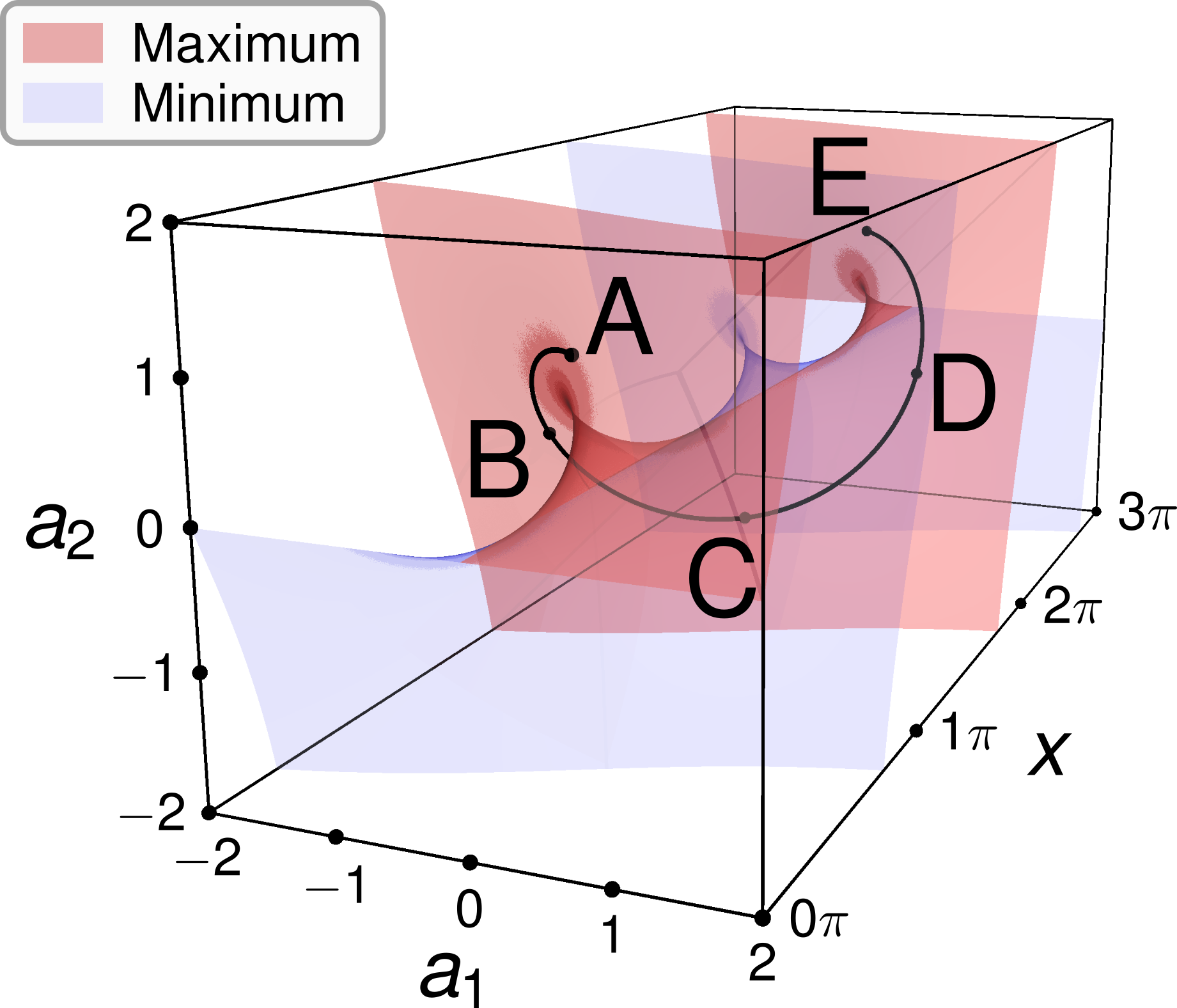}};
        
        \node[black, thick, inner sep=0pt, right=1em and 0em of 3d-view.north east, anchor=north west] (realizations) {%
            \includegraphics[%
                height=4.3cm
            ]{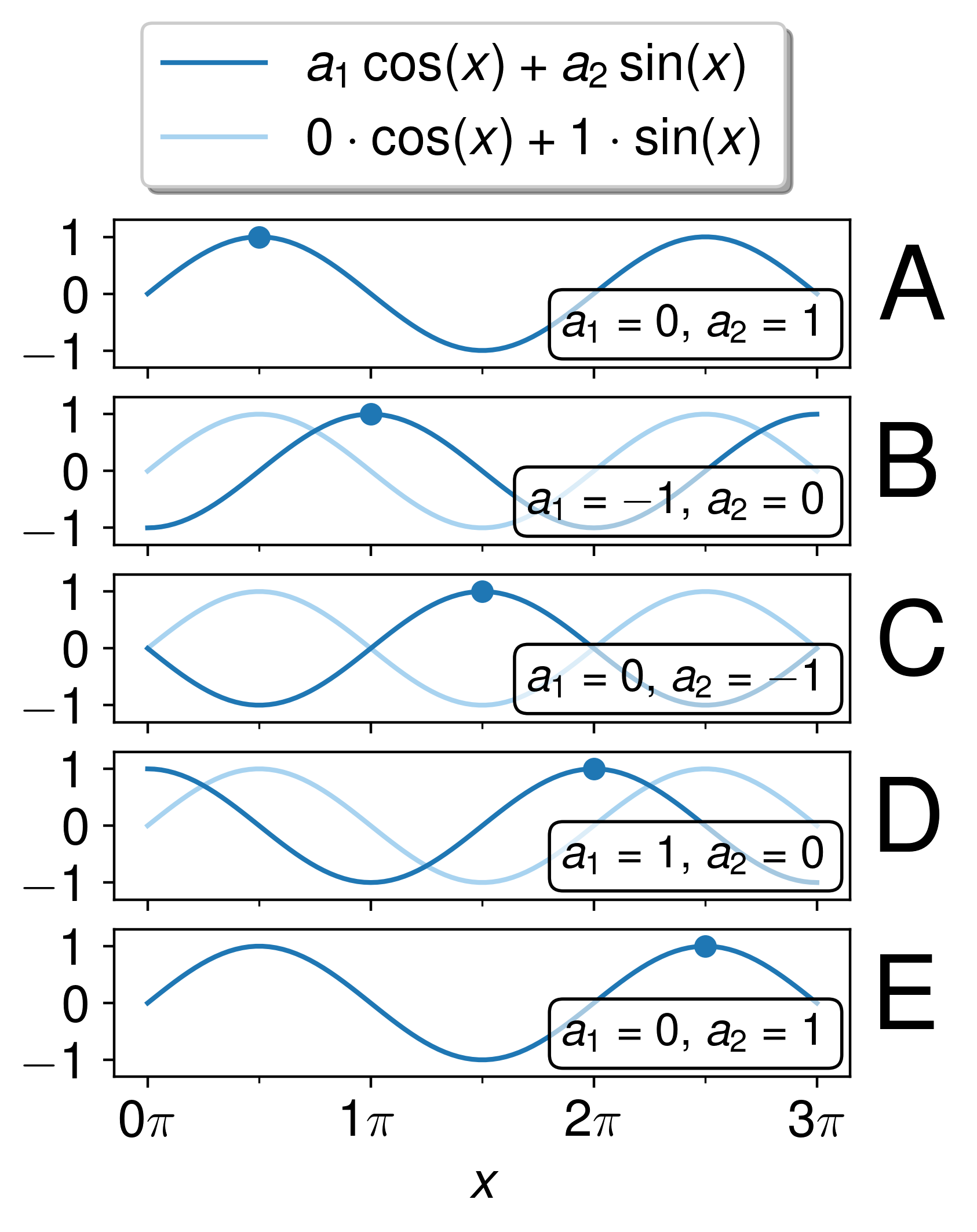}};
    \end{tikzpicture}
  \caption{Left: Part of the critical point set of $\fun_{\vec{\param}}(\gridPt):= \param_1 \cos(\gridPt) + \param_2 \sin(\gridPt)$. A red and a blue component indicate an uncertain maximum and minimum, respectively. Right: Realizations of $\fun_{\vec{\param}}$ for the parameter values at the five points A--E on the uncertain maximum feature a maximum (blue dot) at the respective $x$-coordinate of A--E. The realization at A is drawn in light blue in all plots.
  }
  \label{fig:1D-2D}
\end{figure}

The following example shows that the condition of $\fun_\param$ depending on one parameter is necessary for the claim of \cref{proposition:projectability-to-param-space} to hold.
\vspace{-1em}
\begin{example}
\label{example:counterexample}
Consider the one-dimensional (linear) \emph{two-parameter} family $\fun_{\vec{\param}}(\gridPt) := \param_1 \cos(\gridPt) + \param_2 \sin(\gridPt)$. 
We find its set of critical points to be $\jacobi_{\fun_\vec{\param}} = \{(\gridPt, \vec{\param}) \in \dom \times \paramSpace\mid \param_2 = \cos(\gridPt) = 0 \lor \param_2 = \param_1 \tan(\gridPt)\}$  (see \cref{eq:set-of-crit-pts}). 
A portion of $\jacobi_{\fun_\vec{\param}}$ for $\gridPt \in [0,3\pi]$ is depicted on the left of \cref{fig:1D-2D}.
One can make out two uncertain critical points, a maximum (red) and a minimum (blue), forming a helicoid around the $\gridPt$-axis. The points A and E both belong to the red component and have coordinates $(\pi/2, 0, 1)$ and $(5\pi/2, 0, 1)$; that is, they have different values for $x$, but the same parameter configuration $\vec{\param} = (0,1) \in \paramSpace$, thus violating the claim of \cref{proposition:projectability-to-param-space} that each uncertain critical point can be projected onto $\paramSpace$ without mapping two points to the same parameter value.

The example further illustrates why this is problematic. 
Because A and E belong to the same uncertain maximum, they are connected by a path in $\jacobi_{\fun_\vec{\param}}$---for example, by the black helix with radius one embedded in the uncertain maximum in the figure. Point $\text{A} = (\pi/2, 0, 1)$ corresponds to the maximum at $\gridPt = \pi/2$ of the realization $\fun_{(0,1)}(\gridPt) = \sin(\gridPt)$ depicted on the top right. 
Now, if we follow the helix, passing through the points B--D, the realizations at the respective parameter configurations and therewith the position of the maximum get continuously shifted to the right (see plots for B--D on the right). Finally, if we further follow the helix to point E, we end up at the same parameter configuration $(0,1) \in \paramSpace$ that we started at, thus resulting in the same function (light blue curve). However, at this point, the maximum, which initially was located at $\gridPt=\pi/2$, now moved to $\gridPt = 5\pi/2$, which was the location of another maximum of the function depicted at the top. That is, our definition of uncertain critical points also identifies spatially separated critical points of the same realization as manifestations of the same feature, which, in a critical point analysis on a per-realization basis, would have been detected as different features.

Another challenge arises when we compute the occurrence probability of the uncertain critical point in a certain range of the domain. Consider, for example, the interval $[0, 2\pi] \subset \dom$. The red component representing the uncertain maximum revolves once around the $\gridPt$-axis in this range. The projection of this part of the helicoid onto $\paramSpace$ covers it entirely (i.e., $P_\ucp([0,2\pi]) = \paramSpace$). Thus---irrespective of the distribution of $\vec{\rvParam}$---the probability that the uncertain maximum will occur in this interval, $\prob(\ucp \text{ in } [0,2\pi])$, is $1$ (\cref{eq:probability-of-region-parametric}). Because of the $2\pi$-periodicity of $\fun_\vec{\param}(\gridPt)$, the same is true for any interval $[2\pi k, 2\pi(k+1)], k \in \mathds{Z}$, giving the counterintuitive result that the \emph{same} critical point occurs with certainty in every such interval.
\end{example}

The example shows that even increasing the dimension of the parameter space to two leads to an ambiguous projection of the field's uncertain critical points onto the parameter space and a critical point with spatial uncertainty whose distribution is difficult to interpret.

It is not immediately clear how to circumvent this problem. Ultimately, the issue arises from too many points being identified as manifestations of the same features, resulting in large uncertain critical points that no longer have an overlap free projection onto $\paramSpace$. A first idea might hence be to search for sensible criteria to further partition uncertain critical points to ensure that they do not contain more than one manifestation of a feature in any realization. However, finding such a criterion can be challenging. Looking at the previous example, there is not one singular way to further subdivide, for instance, the red uncertain maximum because any partition could also be performed with an arbitrary phase shift (i.e., rotated along the helicoid), producing a qualitatively equivalent result.

Another option is to accept that the same feature can have two manifestations in the same realization and that the spatial distribution might not be additive. However, this requires visualizations that effectively communicate such dependencies to prevent the viewer from misinterpretations. For example, the viewer should not mistake the spatial distribution of an uncertain critical point for a probability distribution because integrating the spatial density over an area does not necessarily give the probability of the uncertain critical point to occur in this area.
Further research in collaboration with domain experts is required to understand how this, at first glance, counterintuitive result is to be interpreted in application domains and how a visual design can support this process.

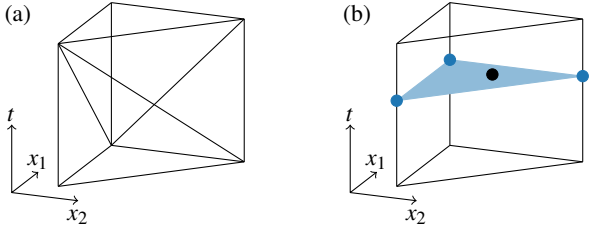
\begin{figure}
    \centering
    \begin{tikzpicture}
        \node[black] (prism){\input{pictures/Interpolation_Schemas/Illustration_Prismatic_Interpolation.tex}};
    \end{tikzpicture}
    \vspace*{-.6em}
    \caption{Interpolations schemes for time-dependent fields. (a) Tetrahedrisation of the space-time domain as suggested by Edelsbrunner \etal{}~\cite{Edelsbrunner2004,Edelsbrunner2008}. Function values within the tetrahedra are interpolated from the values at their corners using barycentric coordinates. (b) Prismatic tessellation suggested by Mascarenhas~\cite{Mascarenhas2006}. To interpolate within the cell, first, the values at the corners of the blue triangle parallel to the domain at the time of interest are interpolated from the values at the adjacent time steps. Then the barycenter of those values is used to retrieve the value at the point of interest (black dot).}
    \label{fig:interpolation}
\end{figure}
\section{Algorithm for Discrete Fields}
\label{sec:algo-discr}
In their work, Liebmann and Scheuermann~\cite{Liebmann2016} analyze the critical points of linear multiparameter families of PL fields on 2D simplicial grids with vertices $\vec{\gridPt_0}, \dots, \vec{\gridPt_\numGridPts}$---that is, fields of the form
\begin{equation}
\label{eq:lin-mul-param-fam-of-fields}
\vec{\fun}_{\vec{\param}} := \vec{\evs_0} + \sum_{i=1}^{\numFields} \param_i \vec{\evs_i},
\end{equation}
with discrete fields $\vec{\evs_0}, \dots, \vec{\evs_\numFields} \in \reals^\numGridPts$ and parameters $\vec{\param} \in \paramSpace = \reals^\numFields$.
To track critical points across different realizations, they introduce singular patches and the concept of patch adjacency. 
In this section, we outline a connection between our generic notion of uncertain critical points, the combinatorial concept of singular patches by Liebmann and Scheuermann~\cite{Liebmann2016}, and previous works on the tracking of critical points through time-dependent PL fields by Edelsbrunner \etal~\cite{Edelsbrunner2008} and Mascarenhas~\cite{Mascarenhas2006}. This will show how a discrete approximation of the uncertain critical points of the specific class of fields in \cref{eq:lin-mul-param-fam-of-fields} can be obtained using these discrete approaches. It also places the combinatorial approach of Liebmann and Scheuermann in a broader mathematical context.

\subsection{Discrete Jacobi Sets}
If the uncertain field in \cref{eq:lin-mul-param-fam-of-fields} were differentiable, we could identify its uncertain critical points by computing $\jacobi_{\fun_\vec{\param}}$ and finding connected components of constant type as outlined in \cref{sec:multiparameter-families}. However, as PL functions are generally not differentiable, the definition of $\jacobi_{\fun_\vec{\param}}$ (\cref{eq:set-of-crit-pts}) is not applicable. In the differentiable case, we found that this set is the Jacobi set of $\fun_\vec{\param}$ and a set of auxiliary functions (\cref{lemma:relation-to-jacobi-set}).
An obvious way to extend $\jacobi_{\fun_\vec{\param}}$ to the PL case is hence to search for a discrete representation $\jacobiDiscr_{\vec{\fun}_\vec{\param}}$ of this Jacobi set.
An approach to this has already been proposed by Edelsbrunner and Harer~\cite{Edelsbrunner2004}. 
Assuming that the input fields are given on the vertices of a triangulation of the ambient space $\dom \times \paramSpace$, they describe an algorithm that extracts a set of $\numFields$-simplices that must belong to $\jacobiDiscr_{\vec{\fun}_\vec{\param}}$. These simplices induce a subcomplex of the triangulation, which they regard as the (discrete) Jacobi set of the PL input fields. 
In theory, we could use this approach to obtain an approximation of $\jacobi_{\fun_\vec{\param}}$ and extract the uncertain critical points from it. In practice, however, triangulating the $(\domainDims + \numFields)$-dimensional ambient space will produce a quickly growing (with $\numFields$) number of simplices resulting in a prohibitive increase in computational cost~\cite{Hughes1996}.

In his thesis on the temporal tracking of Reeb graphs, Mascarenhas~\cite{Mascarenhas2006} discusses further shortcomings of this approach in the analyzed time-dependent (i.e., one-parametric) case and proposes to use a prismatic representation of $\dom \times \reals$ instead, where the fields' values at discrete time steps are given on the vertices of the same triangulation of $\dom$. \Cref{fig:interpolation} illustrates the interpolation schemas used by Edelsbrunner \etal~\cite{Edelsbrunner2004, Edelsbrunner2008} and Mascarenhas~\cite{Mascarenhas2006}. 
\begin{figure}
    \centering
    \begin{tikzpicture}
        \node[black] (prism-triangulated){\input{pictures/Tracking_Mascarenhas/Illustration_Prismatic.tex}};
    \end{tikzpicture}
    \vspace*{-1.4em}
    \caption{Movement of a minimum between two adjacent time steps of a time-dependent field. (a) Part of the prismatic tessellation of $\dom\times \reals$ showing one grid point and its six neighbors at two time steps. (b-d) Configurations of the central grid point's link. Dots with a ``$+$'' mark neighbors with higher, dots with a ``$-$'' neighbors with lower values. Three critical intervals (solid blue lines) mark time ranges where three grid points attain a minimum. Dashed lines mark events where the link of the central grid point changes. 
    }
    \label{fig:tracking-mascarenhas}
\end{figure}
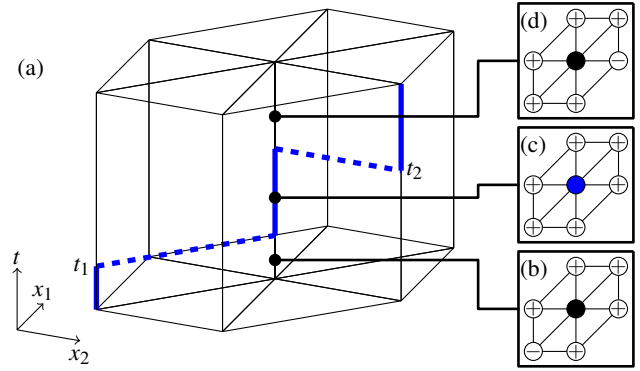
To compute a discrete representation of the Jacobi set using the prismatic interpolation, Mascarenhas examines each temporal edge in the prismatic mesh for time intervals where the grid point is critical (see \cref{fig:tracking-mascarenhas}). 
For this, his method first computes the grid point's lower link at the earlier time step (dots with ``$-$'' in \crefSubFigRef{fig:tracking-mascarenhas}{b}). Then it finds all (sorted) moments in time where a neighboring grid point enters or leaves the lower link. For instance, at $t_1$, the lower left grid point leaves the lower link of the grid point, causing it to become a minimum (\crefSubFigRef{fig:tracking-mascarenhas}{c}) before it becomes a regular point again when at $t_2$ the right neighbor enters the lower link (\crefSubFigRef{fig:tracking-mascarenhas}{d}). This results in a decomposition of the temporal edge into \emph{critical intervals} of constant lower link (solid blue lines), which can be joined by edges connecting two vertices of the grid in the same time slice (dashed lines at $t_1$ and $t_2$) where a critical point travels in between grid points. 
The resulting collection of all those lines is a discrete approximation of the Jacobi set tracing the temporal movement of the field's critical points.

\subsection{Connection to Singular Patches}
In the following, we adapt Mascarenhas' approach for the tracking of critical points through time for the analysis of the critical points of the linear multiparameter family in \cref{eq:lin-mul-param-fam-of-fields}. We will make two adjustments: First, we move from function values at discrete steps of the bounded time dimension to the continuous evolution of function values along a single unbounded parameter dimension. This gives an algorithm for computing $\jacobiDiscr_{\vec{\fun}_\param}$ in the one-parametric case. Then we move on to higher dimensional parameter spaces, where we will see that the notions of singular patches and their adjacency offer a natural extension to the tracking approach by Mascarenhas.

\noindent\textbf{Step 1.} We consider the one-parametric function family $\vec{\fun}_\param = \vec{\evs_0} + \param \vec{\evs_1}$.
To find the events that bound the critical intervals, Mascarenhas only requires the function values at the corners of each prism (see \crefSubFigRef{fig:interpolation}{b}) to be linear functions---in his case, obtained from linear interpolation between the function values at the adjacent time steps.
In our case, the function value at the $i^\text{th}$ grid point is given by a linear function in $\param$: $\fun_{\param;i} = \evs_{0;i} + \param \evs_{1;i}$. Thus, instead of multiple bounded prisms between consecutive time steps, we have only one prism per spatial grid cell, which is unbounded along the parameter dimension. Where Mascarenhas started with the configuration of a grid point's lower link at the earlier time step of the temporal edge, we therefore start with the lower link at $a\to -\infty$. Because, in this limit, $\evs_{0;i}$ becomes negligible, we start with the lower link of $-\evs_{1;i}$ and then trace its change for increasing values of $\param$ to find critical intervals. We then connect the critical intervals of neighboring grid points that share the same change event. Finally, we split any monkey saddles following the approach described in Mascarenhas' work.

\begin{figure}
    \centering
    \includegraphics[width=.95\linewidth]{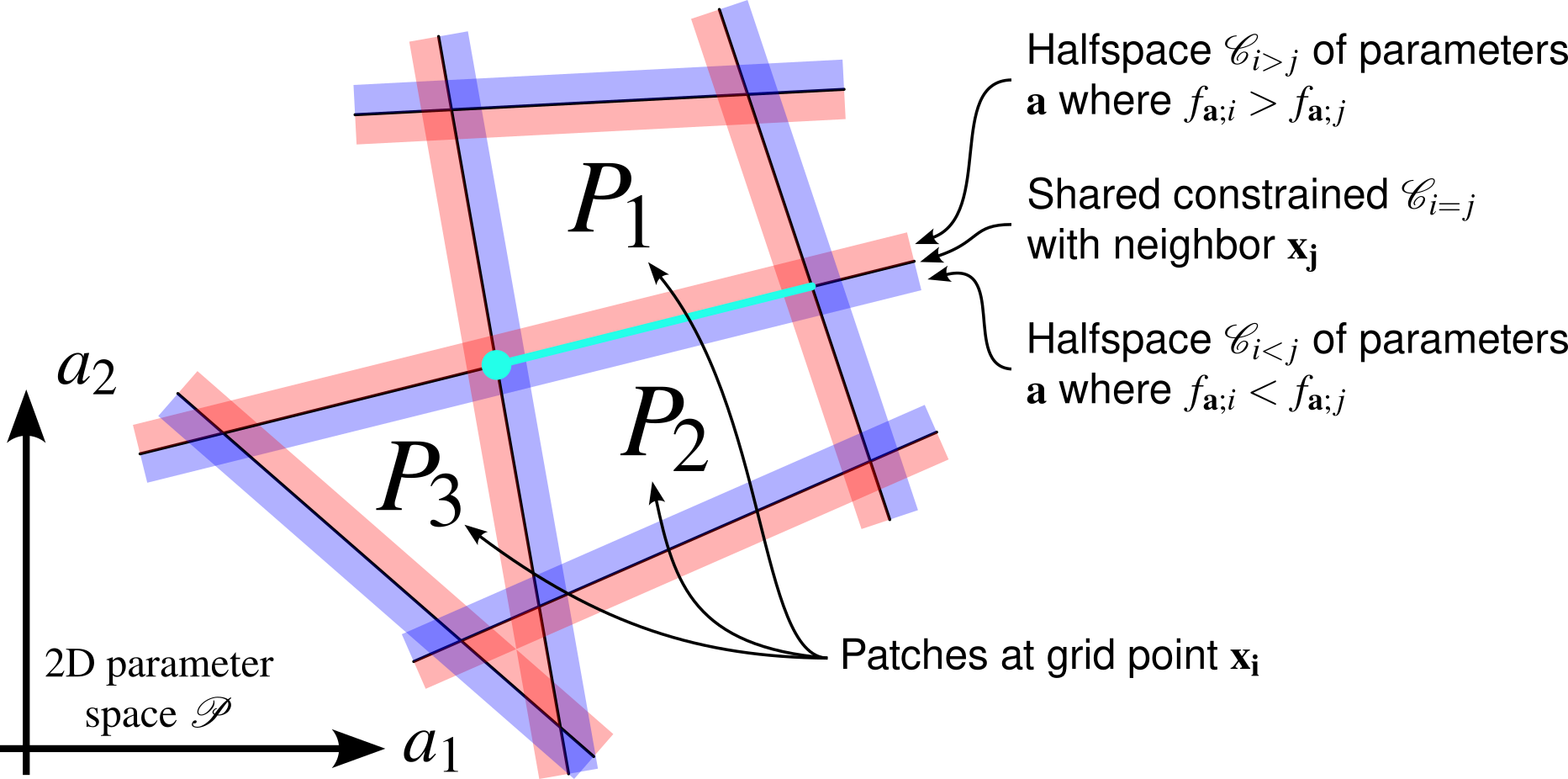}
    \caption{Three patches $P_1$, $P_2$, and $P_3$ of some grid point $\vec{\gridPt_i}$ are obtained from partitioning the 2D parameter space $\paramSpace$ along all hyperplanes where $\vec{\gridPt_i}$ is mapped to the same value as one of its neighbors (black lines). $P_1$ and $P_2$ intersect in a 1D polytope (turquoise line segment) and $P_1$, $P_2$ and $P_3$ intersect in a 0D polytope (turquoise dot).}
    \label{fig:patches}
\end{figure}
\noindent\textbf{Step 2.} Now, we extend this approach to higher dimensions of the parameter space (i.e., $m > 1$). For this, we build on the notion of \emph{singular patches} introduced by Liebmann and Scheuermann~\cite{Liebmann2016}. We briefly recapitulate that concept: Consider two grid points $\vec{\gridPt_i}, \vec{\gridPt_j}$ of $\grid$. 
Because their function values $\fun_{\vec{\param};i}$ and $\fun_{\vec{\param};j}$ depend linearly on $\vec{\param}$, the set of parameters where both grid points are mapped to the same value, $\constraint_{i = j} := \{\vec{\tilde{\param}} \in \paramSpace \mid \fun_{\vec{\tilde{\param}};i} = \fun_{\vec{\tilde{\param}};j}\}$, constitutes a hyperplane in $\paramSpace$ (black lines in \cref{fig:patches}). It partitions $\paramSpace$ in two (open) halfspaces $\constraint_{i < j} := \{\vec{\tilde{\param}} \in \paramSpace \mid \fun_{\vec{\tilde{\param}};i} < \fun_{\vec{\tilde{\param}};j}\}$ and $\constraint_{j < i}$ of parameters for which $\fun_{\vec{\param};i}$ or $\fun_{\vec{\param};j}$ has the lower value, respectively (red and blue regions in \cref{fig:patches}). 
Next, consider a grid point $\vec{\gridPt_i}$ and let $\vec{\gridPt_{n_1}},\dots, \vec{\gridPt_{n_k}}$ its direct neighbors in the grid. The collection of all constraints $\constraint_{i = {n_\ell}}, \ell = 1,\dots, k$ then partitions $\paramSpace$ into open polytopic regions $P^\circ\subset \paramSpace$ of parameters for which $\vec{\gridPt_i}$ has unchanged lower link---that is, $\sgn(\fun_{\vec{\param};i} - \fun_{\vec{\param};n_\ell}) = \text{const}$ for all $\vec{\param} \in P^\circ$ and all neighbors $\vec{\gridPt_{n_\ell}}$ of $\vec{\gridPt_i}$. 
The closures $P:=\overline{P^\circ}$ of those regions are called \emph{patches} of $\vec{\gridPt_i}$ ($P_1$, $P_2$, and $P_3$ in \cref{fig:patches}).
If the structure of $\vec{\gridPt_i}$'s lower link within $P^\circ$ indicates a critical point, the patch is called \emph{singular}.
Two patches of the same or neighboring grid points are called \emph{adjacent} if they intersect in a face of dimension less than $\numFields$---that is, if the polytope described by the constraints of both patches is not empty. This information is stored in a \emph{singular patch graph} whose nodes represent the patches and edges connect nodes of adjacent patches. 

We can use these concepts to extend the approach for computing the Jacobi set of a one-parametric field to fields given by \cref{eq:lin-mul-param-fam-of-fields}: As before, the core idea is to find regions of constant lower link configuration at each grid point (vertical lines in \crefSubFigRef{fig:tracking-mascarenhas}{a}) and then to connect regions that share a change event (horizontal lines in \crefSubFigRef{fig:tracking-mascarenhas}{a}). For each grid point $\vec{\gridPt}$, we obtain regions in $\paramSpace$ of constant lower link by computing its patches. Because each singular patch $P$ lives in $\paramSpace$, we need to lift it to the ambient space $\dom \times \paramSpace$ of the Jacobi set by including $\{\vec{\gridPt}\} \times P$ in $\jacobiDiscr_{\vec{\fun}_\vec{\param}}$. 
Two lifted patches belonging to neighboring grid points have a common change event if they are bounded by some common constraint $\constraint_{i = j}$---that is, if they were adjacent in $\paramSpace$. In this case, the two lifted versions of those patches are connected in $\jacobiDiscr_{\vec{\fun}_\vec{\param}}$ by an $\numFields$-dimensional polytope obtained from extruding their common subset $U$ on the constraint $\constraint_{i = j}$ (turquoise line in \cref{fig:patches}) between the two corresponding grid points: $\{\lambda \vec{\gridPt_i} + (1-\lambda) \vec{\gridPt_j}\mid \lambda \in [0,1]\} \times U$ (dashed line in \crefSubFigRef{fig:tracking-mascarenhas}{a}).
In general, if a patch at $\vec{\gridPt_i}$ shares $k$ constraints, $\constraint_{i=n_1}, \dots, \constraint_{i=n_k}$, with patches at neighboring grid points, $\vec{\gridPt_{n_1}},\dots, \vec{\gridPt_{n_k}}$---that is, $\vec{\gridPt_i}, \vec{\gridPt_{n_1}},\dots, \vec{\gridPt_{n_k}}$ attain the same value in some subset $U$ of the $(\numFields-k)$-dimensional intersection $\constraint_{i=n_1}\cap \dots \cap \constraint_{i=n_k} \subset \paramSpace$ (e.g., turquiose dot in \cref{fig:patches} for $k=2$)---the corresponding patches get connected in $\jacobiDiscr_{\vec{\fun}_\vec{\param}}$ by a polytope of dimension at most $\numFields$ obtained from extruding $U$ along the convex hull of these $k+1$ points: $\operatorname{conv}(\vec{\gridPt_i}, \vec{\gridPt_{n_1}}, \dots, \vec{\gridPt_{n_k}}) \times U$.
Note that the convex hull has dimension at most $\domainDims$, such that in case of $k > \domainDims$, the extruded polytope has dimension less than $\numFields$. In that case, the polytope will already be contained in $\jacobiDiscr_{\vec{\fun}_\vec{\param}}$ as part of an $\numFields$-dimensional polytope resulting from a subset of the grid points having the same value.
In summary, the discrete Jacobi set $\jacobiDiscr_{\vec{\fun}_\vec{\param}}$ consists of two types of $\numFields$-dimensional polytopic regions: singular patches living in some lifted parameter space $\{\vec{\gridPt}\} \times \paramSpace$ (generalizing critical intervals; solid lines in \crefSubFigRef{fig:tracking-mascarenhas}{a}) and polytopes obtained from extruding an $(\numFields-k)$-dimensional region $U \subset \paramSpace$ along a $k$-dimensional convex set in $\dom$ spawned by a set of $k\leq \domainDims$ grid points (generalizing lines connecting critical intervals; dashed lines in \crefSubFigRef{fig:tracking-mascarenhas}{a}).
The singular patch graph can then be seen as an abstract description of this geometric construct.

In theory, the above procedure could be used to compute a geometric representation of $\jacobiDiscr_{\vec{\fun}_\vec{\param}}$ within $\dom\times \paramSpace$. In practice, however, this will typically be a high-dimensional object which cannot be visualized directly. As our original goal was to characterize the spatial uncertainty of critical points and uncertain critical points mainly constitute a mathematical concept to achieve this goal, we can dispense with their explicit computation. Instead, we can compute the spatial projection of the uncertain critical points following the approach by Liebmann and Scheuermann of finding connected components of the singular patch graph while only considering edges connecting singular patches of the same type.
But in order to find the full spatial extend of each uncertain critical point, we do not stop the connected component search prematurely, even if this means merging patches that overlap in parameter space. 

A limitation of the algorithm described by Liebmann and Scheuermann is its computational complexity, which depends on the number of grid neighbors of the vertices in the grid. 
 As this number quickly increases with the dimension of the domain, the extraction of critical points with spatial uncertainty using the approach outlined above might already be too expensive for data sets on 3D domains.

\subsection{Visual Representation}
Liebmann and Scheuermann visualize critical points with spatial uncertainty by drawing an outline around all points that belong to the
same extracted connected component.
Note that because the patches live at the grid points, projecting them onto the domain would result in a discrete set of points. Only additionally projecting the polytopes connecting the patches produces continuous regions of the domain, justifying their visual representation as connected regions by Liebmann and Scheuermann. Moreover, while a singular patch $P$ is an $\numFields$-dimensional polytopic region of the $\numFields$-dimensional parameter space $\paramSpace$, generally resulting in a non-zero integral of the probability density over all $\vec{\param} \in P$, the polytopes connecting the patches stem from extruding a polytope in $\paramSpace$ of dimension less than $\numFields$ in orthogonal directions to $\paramSpace$ and hence cover a null set of the parameter space, resulting in an accumulated probability of zero. That is, the occurrence probabilities for an uncertain critical point are limited to the grid points. A Monte-Carlo sampling approach to estimate those probabilities was described by Liebmann and Scheuermann.
While not immediately clear from the text, Liebmann and Scheuermann seem to depict critical points with spatial uncertainty by interpolating the probabilities in between neighboring vertices belonging to the same extracted connected component. This approach can, of course, be applied analogously to depict the projection of uncertain critical points. Note, however, that because all density information is only given at the grid points, the output of this method---similar to those of the local methods discussed in the introduction---depends on the grid resolution (see \cref{fig:shortcomings-local-methods}). A simple approach to counteract this effect is to normalize the probability of each patch by dividing it by the area of the part of the uncertain critical point's projection onto $\dom$ that falls into the cell surrounding the grid point of that patch.

\subsection{Real-World Example}
In this section, we demonstrate that the interpretational issues discussed in \cref{sec:spatial-distribution,sec:spatial-distribution2} arise not only in artificially constructed cases, such as the one in \cref{example:counterexample}, but also in real-world data sets.

For this, we analyzed the sea level pressure fields over the North Atlantic of the 2000 years long preindustrial control simulation of the Max Plank Institute for Meteorology's Grand Ensemble (MPI-GE)~\cite{Maher2019}. Due to the absence of external forcings, this simulation captures the internal variability of Earth's climate, hence effectively sampling the realization space of sea level pressure fields. To obtain a linear multiparameter representation of this ensemble, we apply an empirical orthogonal function (EOF) analysis~\cite{Lorenz1956} giving us a mean field $\vec{\evs_0}$ and a set of leading EOF modes $\vec{\evs_1},\dots, \vec{\evs_\numFields}$ that we can use in the linear representation in \cref{eq:lin-mul-param-fam-of-fields}. To highlight how quickly the mentioned issues arise, we restricted the reconstruction to the first two EOFs (i.e., $\numFields=2$). For the computation of the patch probabilities, we assumed that the parameter vector $\vec{\param}$ stems from a standard bivariate normal distribution (i.e., $\vec{\rvParam} \sim \mathcal{N}(\vec{0}, \vec{I})$). A Q-Q plot supporting this assumption is contained in the supplement.

\crefSubFigRef{fig:NAO-EAP}{a} depicts the spatial extent (i.e., the projection onto the domain) of the uncertain maximum with the highest probability. In total, the probability for the maximum to occur anywhere in the outlined region is more than 99\%. However, if we split that region along the dashed black line, we find that the \emph{same} maximum occurs with a probability of about 68\% in the left part and a probability of 98\% in the right part. That is, the probability of the maximum to occur in the entire region is not split up on both parts as might be intuitive. This is again due to the \emph{same} maximum occurring at different locations in the same realization. To highlight this, \crefSubFigRef{fig:NAO-EAP}{b} shows two maximum patches that belong to the uncertain maximum for the grid points marked $A$ and $B$, respectively. The integrated probability density over patch $A$ is 47\% and over patch $B$ 39\%. The two patches have a large intersection in $\paramSpace$ containing all realizations where the \emph{same} maximum occurs \emph{both} at $A$ and $B$. The collective probability for all those realizations is 36\%.

Further discussion on this real-world application, including an analysis of the case $m=1$, is given in the supplemental material.

\begin{figure}
    \centering
    \begin{tikzpicture}
        \pdfpxdimen=\dimexpr 1000mm/4724\relax
        \node[draw, black, thick, inner sep=0pt] (maximum) {%
            \includegraphics[trim={1100px 260px 840px 750px},clip,height=3.7cm]{%
            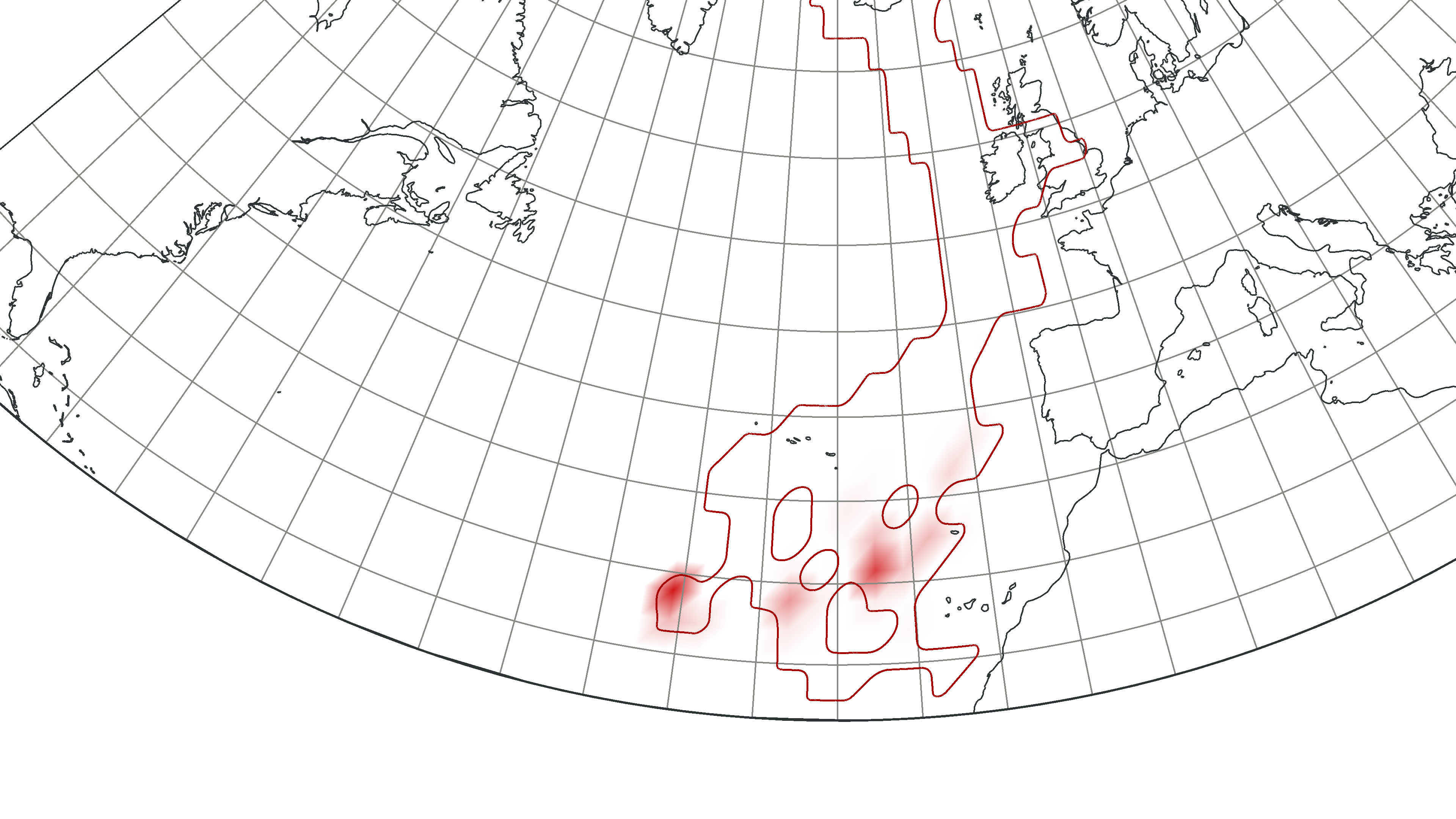}};
        \node[black, thick, inner sep=0pt, left=0.3em of maximum] (colorbar) {%
            \includegraphics[trim={3670px 560px 0px 560px},clip,height=3.7cm]{%
            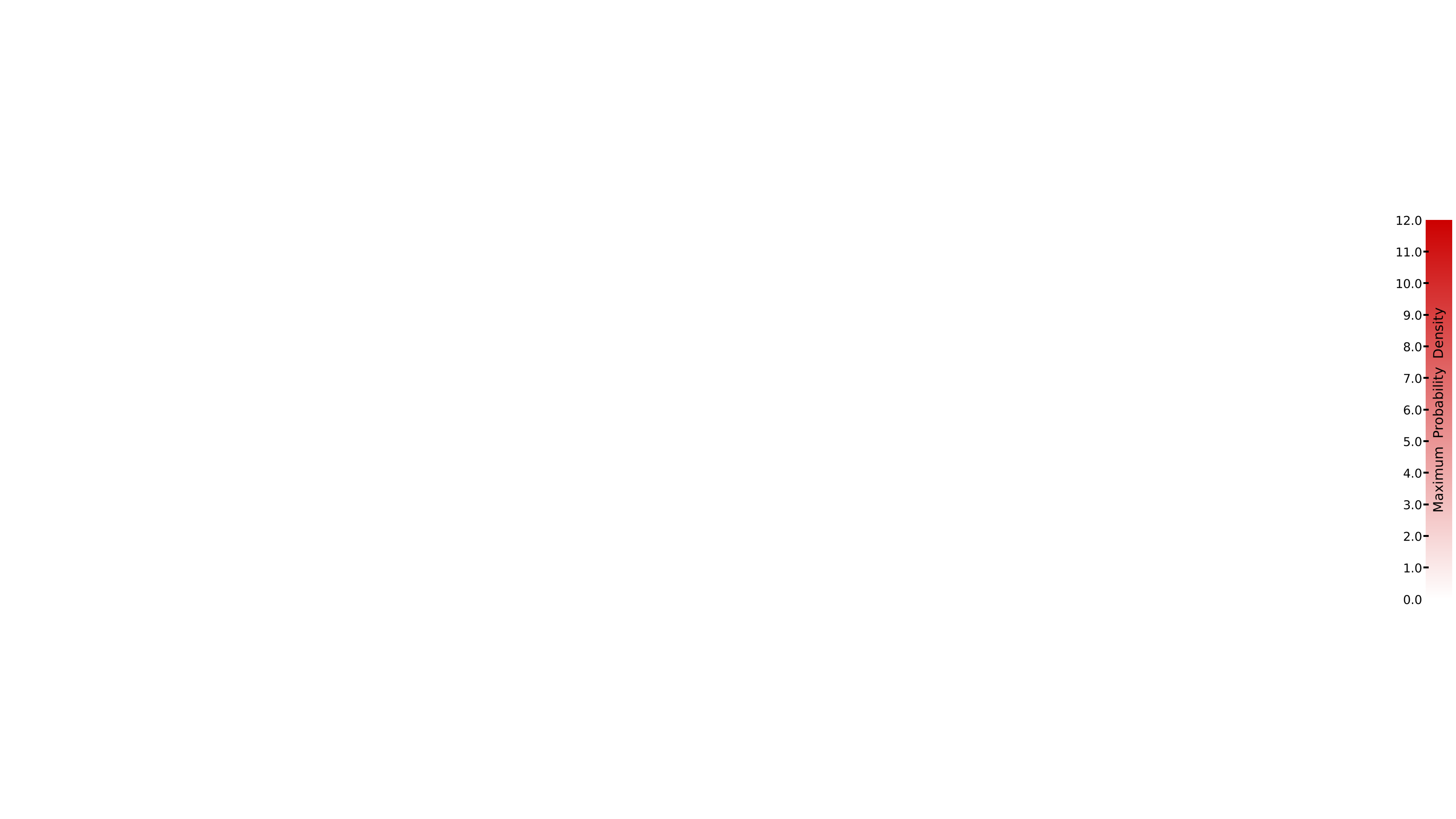}};
        \node[black, thick, inner sep=0pt, right=0.3em of maximum] (patches) {%
            \includegraphics[height=3.7cm]{%
            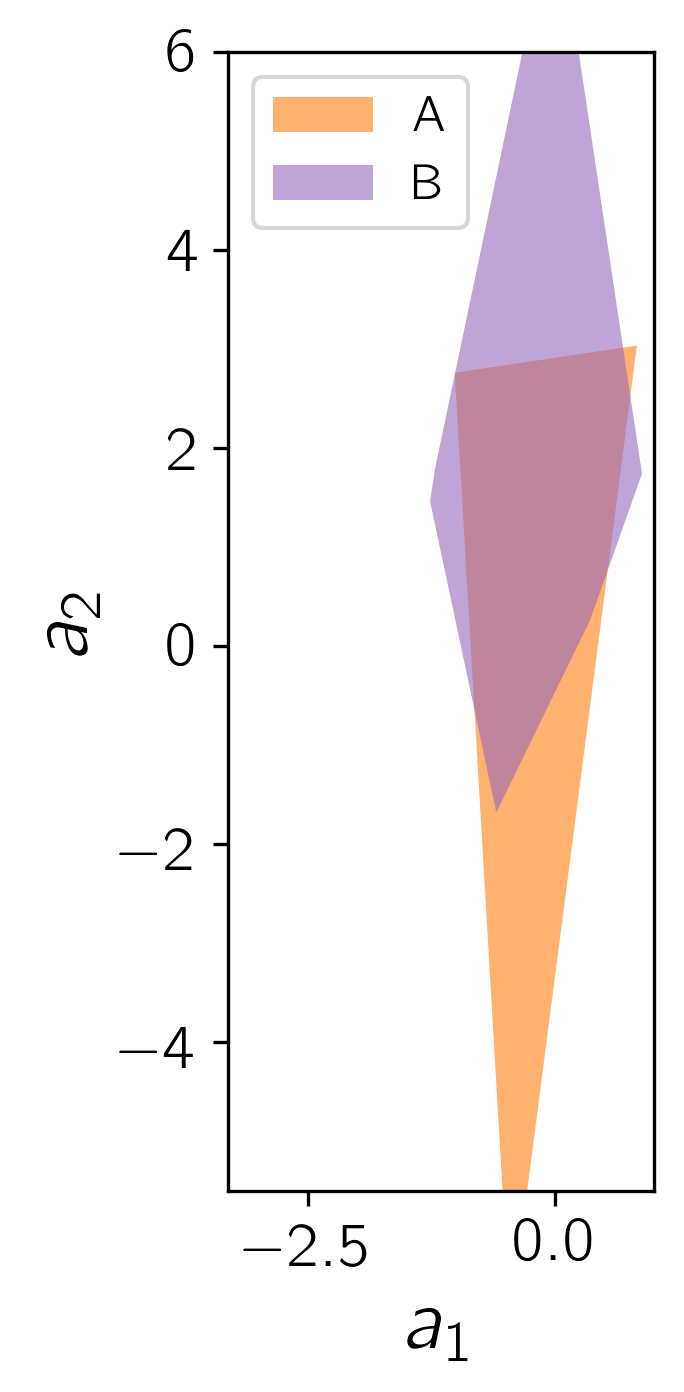}};
            
        \coordinate (ul) at (maximum.north west);
        \coordinate (ur) at (maximum.north east);
        \coordinate (ll) at (maximum.south west);
        \coordinate (lr) at (maximum.south east);
        
        \draw[black, thick, dashed] ($(ul)+(10em,-3.0em)$)
                              -- ($(ul)+(10em,-6.9em)$) 
                              -- ($(ul)+(11.2em,-8.6em)$)
                              -- ($(ul)+(11.2em,-11.5em)$);
                              
        \node[inner sep=0pt] at ($(maximum.center) + (-1em, -4em)$) (label_A) {A};
        \node[inner sep=0pt] at ($(maximum.center) + (4em, -.2em)$) (label_B) {B};
        
        \draw[black, thick, ->] (label_A.north west) -- ($(maximum.center) + (-2.5em, -2.7em)$);
        \draw[black, thick, ->] (label_B.south west) -- ($(maximum.center) + (3em, -1.5em)$);

        \node[inner sep=0pt, below right=0.5em and 0.5em of maximum.north west, anchor=north west] (label_a) {(a)};
        \node[inner sep=0pt, below right=0.5em and 0.1em of patches.north west, anchor=north west] (label_b) {(b)};
    \end{tikzpicture}
    \caption{(a) Uncertain maximum with the highest probability in the two-parametric approximation of sea level pressure over the North Atlantic. (b) Overlapping maximum patches of the uncertain maximum at the marked grid points in (a).}
    \label{fig:NAO-EAP}
\end{figure}
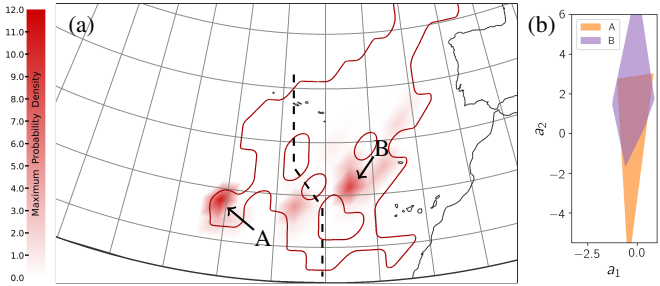

%% file: pictures/Interpolation_Schemas/Illustration_Prismatic_Interpolation.tex
\begin{tikzpicture}[line join=round]
\definecolor{pyplotblue}{HTML}{1F77B4}\draw(5.825,.628)--(5.825,1.767);
\draw(1.325,.628)--(.618,.082);
\draw(3.091,.409)--(1.325,.628);
\draw(1.325,.628)--(1.325,2.527);
\draw(1.325,.628)--(.618,1.981);
\draw(1.325,.628)--(3.091,2.309);
\draw(5.825,.628)--(5.118,.082);
\draw(7.591,.409)--(5.825,.628);
\draw(5.825,1.767)--(5.825,2.527);
\filldraw[color=pyplotblue!50](5.825,1.767)--(5.118,1.222)--(7.591,1.549)--cycle;
\filldraw[color=pyplotblue](5.825,1.767) circle (.075);
\draw(5.825,2.527)--(5.118,1.981);
\draw(7.591,2.309)--(5.825,2.527);
\draw(3.091,2.309)--(1.325,2.527);
\draw(1.325,2.527)--(.618,1.981);
\draw(7.591,.409)--(7.591,2.309);
\draw(5.118,.082)--(7.591,.409);
\draw(3.091,.409)--(.618,1.981);
\draw(3.091,.409)--(3.091,2.309);
\draw(.618,.082)--(3.091,.409);
\filldraw[color=black](6.39,1.571) circle (.075);
\filldraw[color=pyplotblue](7.591,1.549) circle (.075);
\draw[arrows=<->](4.853,.273)--(4.5,0)--(4.5,.904);
\draw[arrows=<->](.353,.273)--(0,0)--(0,.904);
\draw(.618,1.981)--(3.091,2.309);
\draw(5.118,1.981)--(7.591,2.309);
\draw(5.118,.082)--(5.118,1.981);
\draw(.618,.082)--(.618,1.981);
\draw[arrows=->](0,0)--(.883,-.109);
\draw[arrows=->](4.5,0)--(5.383,-.109);
\filldraw[color=pyplotblue](5.118,1.222) circle (.075);
\node[inner sep=1pt, anchor=south] at (.353,.273) {$x_1$};
    	     \node[inner sep=1pt, anchor=south] at (0,.904) {$t$};
	     \node[inner sep=3pt, anchor=north] at (.883,-.109) {$x_2$};\node[inner sep=0pt, anchor=north west] at (-.1,2.5) {(a)};\node[inner sep=1pt, anchor=south] at (4.853,.273) {$x_1$};
    	     \node[inner sep=1pt, anchor=south] at (4.5,.904) {$t$};
	     \node[inner sep=3pt, anchor=north] at (5.383,-.109) {$x_2$};\node[inner sep=0pt, anchor=north west] at (4.4,2.5) {(b)};\end{tikzpicture}

%% file: pictures/Tracking_Mascarenhas/Illustration_Prismatic.tex
\begin{tikzpicture}[line join=round]
\draw(.698,.702)--(.698,3.588);
\draw(.698,.702)--(0,0);
\draw(2.375,.41)--(.698,.702);
\draw(.698,.702)--(.698,3.588);
\draw(.698,.702)--(-1.676,.293);
\draw(0,0)--(.698,.702);
\draw(1.676,-.293)--(2.375,.41);
\draw(2.375,.41)--(2.375,3.295);
\draw(2.375,.41)--(0,0);
\draw(0,0)--(2.375,.41);
\draw(2.375,.41)--(2.375,3.295);
\draw(-1.676,.293)--(-2.375,-.41);
\draw(-1.676,.293)--(-1.676,3.178);
\draw(-1.676,.293)--(0,0);
\draw(0,0)--(-1.676,.293);
\draw(-1.676,.293)--(-1.676,3.178);
\draw(2.375,3.295)--(.698,3.588);
\draw(.698,3.588)--(0,2.885);
\draw(.698,3.588)--(-1.676,3.178);
\draw(0,2.885)--(.698,3.588);
\draw(0,0)--(0,2.885);
\draw(0,0)--(-.698,-.702);
\draw(1.676,-.293)--(0,0);
\draw(0,0)--(0,2.885);
\draw(0,0)--(1.676,-.293);
\draw(0,0)--(0,2.885);
\draw(0,0)--(0,2.885);
\draw(-.698,-.702)--(0,0);
\draw(0,0)--(-2.375,-.41);
\draw(0,0)--(0,2.885);
\draw(-2.375,-.41)--(0,0);
\draw(0,0)--(0,2.885);
\draw[color=blue,line width=2](0,.577)--(0,1.731);
\draw[color=blue,line width=2,dashed](-2.375,.167)--(0,.577);
\draw(2.375,3.295)--(0,2.885);
\draw(1.676,2.593)--(2.375,3.295);
\draw(0,2.885)--(2.375,3.295);
\draw(1.676,-.293)--(1.676,2.593);
\draw(-.698,-.702)--(1.676,-.293);
\draw(1.676,-.293)--(1.676,2.593);
\draw(-1.676,3.178)--(0,2.885);
\draw(-1.676,3.178)--(-2.375,2.476);
\draw(0,2.885)--(-1.676,3.178);
\draw[arrows=<->](-3.08,-.328)--(-3.429,-.679)--(-3.429,.145);
\draw[color=blue,line width=2,dashed](0,1.731)--(1.676,1.439);
\draw(-2.375,-.41)--(-.698,-.702);
\draw(-2.375,-.41)--(-2.375,2.476);
\draw[color=blue,line width=2](-2.375,-.41)--(-2.375,.167);
\draw(-2.375,-.41)--(-2.375,2.476);
\draw(0,2.885)--(-2.375,2.476);
\draw(-.698,2.183)--(0,2.885);
\draw(0,2.885)--(1.676,2.593);
\draw(1.676,2.593)--(0,2.885);
\draw(0,2.885)--(-.698,2.183);
\draw(-2.375,2.476)--(0,2.885);
\draw[color=blue,line width=2](1.676,1.439)--(1.676,2.593);
\draw[arrows=->](-3.429,-.679)--(-2.591,-.825);
\draw(-.698,-.702)--(-.698,2.183);
\draw(-.698,-.702)--(-.698,2.183);
\draw(-.698,2.183)--(1.676,2.593);
\draw(-2.375,2.476)--(-.698,2.183);
\draw[line width=1](4.764,.364)--(3.236,.364)--(3.236,-1.164)--(4.764,-1.164)--(4.764,.364);
\filldraw[fill=white](4.566,.166)--(4,.166)--(3.434,-.4)--(3.434,-.966)--(4,-.966)--(4.566,-.4)--cycle;
\draw(4,-.4)--(4.566,.166);
\draw(4,-.4)--(4,.166);
\draw(4,-.4)--(3.434,-.4);
\draw(4,-.4)--(3.434,-.966);
\draw(4,-.4)--(4,-.966);
\draw(4,-.4)--(4.566,-.4);
\filldraw[color=black](4,-.4) circle (.12);
\filldraw[color=black,fill=white](4.566,.166) circle (.12);
\filldraw[color=black,fill=white](4,.166) circle (.12);
\filldraw[color=black,fill=white](3.434,-.4) circle (.12);
\filldraw[color=black,fill=white](3.434,-.966) circle (.12);
\filldraw[color=black,fill=white](4,-.966) circle (.12);
\filldraw[color=black,fill=white](4.566,-.4) circle (.12);
\draw[line width=1](4.764,2.014)--(3.236,2.014)--(3.236,.486)--(4.764,.486)--(4.764,2.014);
\filldraw[fill=white](4.566,1.816)--(4,1.816)--(3.434,1.25)--(3.434,.684)--(4,.684)--(4.566,1.25)--cycle;
\draw(4,1.25)--(4.566,1.816);
\draw(4,1.25)--(4,1.816);
\draw(4,1.25)--(3.434,1.25);
\draw(4,1.25)--(3.434,.684);
\draw(4,1.25)--(4,.684);
\draw(4,1.25)--(4.566,1.25);
\filldraw[color=black,fill=blue](4,1.25) circle (.12);
\filldraw[color=black,fill=white](4.566,1.816) circle (.12);
\filldraw[color=black,fill=white](4,1.816) circle (.12);
\filldraw[color=black,fill=white](3.434,1.25) circle (.12);
\filldraw[color=black,fill=white](3.434,.684) circle (.12);
\filldraw[color=black,fill=white](4,.684) circle (.12);
\filldraw[color=black,fill=white](4.566,1.25) circle (.12);
\draw[line width=1](4.764,3.664)--(3.236,3.664)--(3.236,2.136)--(4.764,2.136)--(4.764,3.664);
\filldraw[fill=white](4.566,3.466)--(4,3.466)--(3.434,2.9)--(3.434,2.334)--(4,2.334)--(4.566,2.9)--cycle;
\draw(4,2.9)--(4.566,3.466);
\draw(4,2.9)--(4,3.466);
\draw(4,2.9)--(3.434,2.9);
\draw(4,2.9)--(3.434,2.334);
\draw(4,2.9)--(4,2.334);
\draw(4,2.9)--(4.566,2.9);
\filldraw[color=black](4,2.9) circle (.12);
\filldraw[color=black,fill=white](4.566,3.466) circle (.12);
\filldraw[color=black,fill=white](4,3.466) circle (.12);
\filldraw[color=black,fill=white](3.434,2.9) circle (.12);
\filldraw[color=black,fill=white](3.434,2.334) circle (.12);
\filldraw[color=black,fill=white](4,2.334) circle (.12);
\filldraw[color=black,fill=white](4.566,2.9) circle (.12);
\filldraw[color=black](0,.25) circle (.075);
\draw[line width=1](0,.25)--(2.7,.25)--(2.7,-.4)--(3.25,-.4);
\filldraw[color=black](0,1.08) circle (.075);
\draw[line width=1](0,1.08)--(2.7,1.08)--(2.7,1.25)--(3.25,1.25);
\filldraw[color=black](0,2.16) circle (.075);
\draw[line width=1](0,2.16)--(2.7,2.16)--(2.7,2.9)--(3.25,2.9);
\node[inner sep=1pt, anchor=south] at (-3.08,-.328) {$x_1$};
    	     \node[inner sep=1pt, anchor=south] at (-3.429,.145) {$t$};
	     \node[inner sep=3pt, anchor=north] at (-2.591,-.825) {$x_2$};\node[inner sep=2pt, anchor=east] at (-2.375,.167) {$t_1$};\node[inner sep=2pt, anchor=west] at (1.676,1.439) {$t_2$};\node[inner sep=2pt, anchor=north west] at (-3.5,3) {(a)};\node[inner sep=0pt, anchor=center] at (4.566,.166) {\scriptsize{$+$}};\node[inner sep=0pt, anchor=center] at (4,.166) {\scriptsize{$+$}};\node[inner sep=0pt, anchor=center] at (3.434,-.4) {\scriptsize{$+$}};\node[inner sep=0pt, anchor=center] at (3.434,-.966) {\scriptsize{$-$}};\node[inner sep=0pt, anchor=center] at (4,-.966) {\scriptsize{$+$}};\node[inner sep=0pt, anchor=center] at (4.566,-.4) {\scriptsize{$+$}};\node[inner sep=0pt, anchor=north west] at (3.25,.265) {(b)};\node[inner sep=0pt, anchor=center] at (4.566,1.816) {\scriptsize{$+$}};\node[inner sep=0pt, anchor=center] at (4,1.816) {\scriptsize{$+$}};\node[inner sep=0pt, anchor=center] at (3.434,1.25) {\scriptsize{$+$}};\node[inner sep=0pt, anchor=center] at (3.434,.684) {\scriptsize{$+$}};\node[inner sep=0pt, anchor=center] at (4,.684) {\scriptsize{$+$}};\node[inner sep=0pt, anchor=center] at (4.566,1.25) {\scriptsize{$+$}};\node[inner sep=0pt, anchor=north west] at (3.25,1.915) {(c)};\node[inner sep=0pt, anchor=center] at (4.566,3.466) {\scriptsize{$+$}};\node[inner sep=0pt, anchor=center] at (4,3.466) {\scriptsize{$+$}};\node[inner sep=0pt, anchor=center] at (3.434,2.9) {\scriptsize{$+$}};\node[inner sep=0pt, anchor=center] at (3.434,2.334) {\scriptsize{$+$}};\node[inner sep=0pt, anchor=center] at (4,2.334) {\scriptsize{$+$}};\node[inner sep=0pt, anchor=center] at (4.566,2.9) {\scriptsize{$-$}};\node[inner sep=0pt, anchor=north west] at (3.25,3.565) {(d)};\end{tikzpicture}

%% file: sections/Conclusion.tex
\section{Conclusion \& Future Work}
\label{sec:conclusion}
In this paper, we studied the question of how to extend the concept of critical points from deterministic real-valued functions to real-valued random fields.
The core idea is the novel concept of uncertain critical points, which group all function realization's critical points that can be continuously traced without change in type while moving through the realization space.
From these sets, we can determine the region where an uncertain critical point can manifest itself and describe fundamental probability events, such as an uncertain critical point having a manifestation in a specific area. For this, no specific distribution of the analyzed fields is assumed.
In the case of realization spaces with a finite parameterization, we showed that uncertain critical points are related to the Jacobi set~\cite{Edelsbrunner2004} of the input field and a set of auxiliary functions by extending results for time-dependent tracking of critical points \cite{Edelsbrunner2008} to multiparameter families of functions. 
We also showed how this connection can be used to approximate the uncertain critical points of discrete random scalar fields represented by probabilistic linear combinations of PL functions by combining prior work on the temporal tracking of critical points in PL fields \cite{Edelsbrunner2008, Mascarenhas2006}  with the concept of singular patches \cite{Liebmann2016}.

Our work was partly motivated by wanting to understand the reluctance of Liebmann and Scheuermann~\cite{Liebmann2016} to merge singular patches in some situations to avoid a certain phenomenon: that uncertain critical points can have more than one manifestation in any realization.
Our investigation showed that the phenomenon is not specific to their algorithm or the class of functions they considered but can arise in the most general mathematical setting.
We proved that the phenomenon cannot arise for one-parameter families of fields and provided an analytic example showing that the phenomenon can occur already for fields depending on two parameters.
We demonstrated that the phenomenon also occurs in a two-parameter real-world example, indicating that it is not purely academic and highlighting the difficulties this entails for communicating the spatial uncertainty of critical points in day-to-day applications.

It is certainly counterintuitive for an uncertain critical point to have more than one manifestation---that is, for the same critical point to occur in more than one place in the same realization. 
On top, this produces counterintuitive results when trying to compute the spatial distribution of an uncertain critical point, leading, for instance, to statements such as ``the critical point is guaranteed to occur in region $A$, but is also guaranteed to occur in another disjoint region $B$.''
How the ensuing difficulty in interpreting the resulting images will impact the practice of domain experts remains to be studied.
If one wishes to avoid this difficulty in interpretation altogether, it is unclear how to proceed.
The phenomenon would become rarer if uncertain critical points could be subdivided further, preferably without giving up the intuitively correct ``if a critical point can be continuously traced to another, they should be considered manifestations of the same feature.''
However, the example we presented (\cref{example:counterexample}) has enough symmetries to indicate that finding such a general, non-arbitrary rule is difficult, and we offer the example as a benchmark instance for this question.

The observed interpretational challenges might also effect the clarity of visualizations that use local approaches to estimate occurrence probabilities~\cite{Vietinghoff2022,Vietinghoff2022a,Guenther2014}. Even if they do not explicitly outline regions of critical points that describe the same feature, viewers might still be tempted to read information on the spatial distribution of a critical point if they see an isolated region of likely locations for critical points. Especially the proposed visual encoding for the spatial uncertainty of mandatory critical points~\cite{Guenther2014} produces color-coded regions similar to the one in \cref{fig:NAO-EAP}, but the paper does not consider the entailing interpretational difficulties that we revealed with our general analysis.
It would be interesting to further investigate if the outputs of those local approaches cause misconceptions by the viewer by means of a user study.

Aside from this interpretational challenge, there are multiple other research directions. \Cref{lemma:relation-to-jacobi-set} shows that, in the parametric case, uncertain critical points are part of the Jacobi set of the uncertain scalar field and a set of auxiliary functions. In the past, Jacobi sets have been found to be sensitive to noise in the data, and different simplification schemas have been proposed (e.g., by Bhatia \etal{}~\cite{Bhatia2015}). However, a detailed analysis of the effects of such simplifications on the uncertain critical points presented here is beyond the scope of this paper and was left for future work. 
In this work, we identified critical points in the realizations of an uncertain scalar field that can be considered the same feature, but did not consider the relationship between different uncertain critical points (e.g., in analogy to the one described by a deterministic field's contour tree).
Mascarenhas~\cite{Mascarenhas2006} described an algorithm for tracking contour trees through time using the discrete approximation of the Jacobi set of a time-dependent field and an auxiliary function. Since the algorithm described in \cref{sec:algo-discr} extends one of his core ideas to multiparameter families of fields, a next step might be to develop approaches for tracking contour trees across the realizations of an uncertain field. However, this is bound to suffer from the same problem outlined in this work.
Finally, it might be interesting to investigate the use of higher-order (e.g., bicubic) interpolants over the domain to extract uncertain critical points using the smooth definition of the Jacobi set instead of the linear approximation described in \cref{sec:algo-discr} to mitigate artifacts such as overlapping uncertain critical points at some grid points.